\documentclass[10pt,leqno]{amsart}
\usepackage{float}
\usepackage{graphicx}
\usepackage{indentfirst,csquotes}

\usepackage{amssymb,amsthm,amsmath}
\usepackage{xcolor,paralist,titlesec,fancyhdr,etoolbox}
\usepackage[colorlinks=true, citecolor=red,linkcolor=black]{hyperref}
\makeatletter
\providecommand*\@dotsep{4.5}
\renewcommand{\l@section}{\@dottedtocline{1}{0.0em}{2.5em}}
\renewcommand{\l@subsection}{\@dottedtocline{2}{3.5em}{2.5em}}
\makeatother

\usepackage{algorithm}
\usepackage{algpseudocode}

\newtheorem{theorem}{Theorem}[]
\newtheorem{definition}[theorem]{Definition}

\newtheorem{proposition}[theorem]{Proposition}
\newtheorem{corollary}[theorem]{Corollary}

\titleformat{\section}{\normalfont\large\bfseries\centering}{\thesection.}{1em}{\MakeUppercase}
\titleformat{\subsection}{\normalfont\normalsize\bfseries}{\thesubsection.}{1em}{}

\newcommand{\ip}[2]{\left\langle\,#1\,|\,#2\,\right\rangle} 
\newcommand{\ket}[1]{|#1\rangle} 
\newcommand{\kb}[2]{|#1\rangle\langle#2|} 
\DeclareMathOperator{\Tr}{Tr}

\begin{document}
\title{Calculating the Projective Norm of higher-order tensors using a
gradient descent algorithm} 
\author[Aaditya Rudra, Maria Anastasia Jivulescu]{Aaditya Rudra, Maria Anastasia Jivulescu}
\address{(AR) Department of Electrical Engineering, Visvesvaraya National Institute of Technology, South Ambazari Road, Nagpur, Maharashtra, 440010, India}
\email{adityarudra02@gmail.com}
\address{(MAJ) Department of Mathematics, Politehnica University of Timi\c soara, Victoriei Square 2, 300006 Timi\c soara, Romania.}
\email{maria.jivulescu@upt.ro}
\date{\today}

\let\thefootnote\relax

\begin{abstract}
Projective Norms are a class of tensor norms that map on the input and output spaces. These norms are useful for providing a measure of entanglement. Calculating the projective norms is an NP-hard problem, which creates challenges in computing due to the complexity of the exponentially growing parameter space for higher-order tensors. We develop a novel gradient descent algorithm to estimate the projective norm of higher-order tensors. The algorithm guarantees convergence to a minimum nuclear rank decomposition of our given tensor. We extend our algorithm to symmetric tensors and density matrices. We demonstrate the performance of our algorithm by computing the nuclear rank and the projective norm for both pure and mixed states and provide numerical evidence for the same.
\end{abstract} 

\maketitle
\tableofcontents

\hypersetup{linkcolor=blue, filecolor=black, urlcolor=black}

\section{Introduction}
Quantum information science brings new ideas in computing and information processing by exploiting the properties of quantum mechanics, such as entanglement and superposition \cite{NC10}. Entanglement is a feature and a resource of the quantum world and  hence, the interest in efficient methods to characterize  and detect it. 
One way of detecting if a quantum state is entangled is to compute its projective norm and check if its value is larger than 1 \cite{R2000}. Nevertheless, the computation of the tensor norm is NP-hard \cite{HL09}, and therefore the need for alternative algorithms for computing it \cite{DFLW17}.

In this paper a novel gradient descent-based algorithm  for computing the projective tensor norm is introduced. The algorithm guarantees the convergence to global minimum norm values, i.e. projective tensor norm, and works for both pure states (rank one density matrices or m-tensors of Hilbert-Schmidt norm equal to 1), and for mixed states (statistical mixtures of pure states or 2m-tensors, usually called density tensors \cite{DFLW17}). Separable quantum states correspond to probabilistic mixtures of product states, and therefore separable density tensors are generalizations
of product states. An entangled quantum state  means that it cannot be written as a product state, i.e. a rank one tensor. Hence,  entangled quantum states are described by density tensors which are not
separable.
Various works have related the detection of entangled multipartite quantum states by the computation of the projective tensor norm associated to the corresponding density tensors \cite{JLN20}. 
For example, the injective tensor norm is related, for pure states, to the geometric measure  entanglement \cite{FLN24}, whereas the projective tensor norm gives a necessary and sufficient condition to test the separability of a quantum states, i.e. to check if the projective norm of the tensor is equal to 1 \cite{R2000}. Relaxations of this result have been introduced in \cite{JLN20} where,  based on linear contraction, called entanglement testers, a unified approach of the entanglement criteria theory based on projective tensor norm is presented.

This paper focuses on developing a new gradient descent algorithm to compute the projective tensor norm of high-order tensors. The paper is organized as follows: Section\eqref{S:PN-def} recalls the main mathematical tools for tensor norms, such as the definitions of projective and injective tensor norms and their mathematical properties.  Section\eqref{S:Alg} explains the existing algorithm for computing the projective norm, as well as the new method. Section\eqref{S:computation} presents the performance of the algorithm, both for pure and mixed states, by comparing them to the analytical/numerical results for some classes of high-order tensors. 

\section{The projective norm and its application in quantum information theory}
\label{S:PN-def}
In this section we recall basic definitions and facts about different natural norms that can be introduced on the algebraic tensor product of finite dimensional Banach spaces. We will focus mostly on the projective norm and its application to detect entangled states. 
We start by defining the projective and the injective tensor norms for (finite dimensional) Banach spaces $V_i, i=1,2,\ldots ,m$. 

\begin{definition}
Given $m$ finite-dimensional Banach spaces $V_1,\ldots,V_m$, with their respective norms $\|\cdot\|_{V_i}$ and $\mathcal{T} \in V_1 \otimes \cdots\otimes V_m$, the \emph{projective (nuclear) tensor norm} of the tensor $\mathcal{T} $ is
\begin{equation}
\label{eq:def-projective-norm}\|\mathcal{T}\|_\pi := \inf \left\{ \sum\limits_{k=1}^{r_N} \|x_k^1\|_{V_1} \cdots \|x_k^m\|_{V_m} \, : \, r_N \in \mathbb N, \ x_k^i\in V_i,\mathcal{T} = \sum_{k=1}^{r_N} x_k^1 \otimes \cdots \otimes x_k^m \right\}.
\end{equation}
Here, the infimum is taken over all the decompositions of the m-tensor $\mathcal{T}= \sum\limits_{k=1}^{r_N} x_k^1 \otimes \cdots \otimes x_k^m,$ where $r_N$ is a finite, but arbitrary integer. The Banach space induced by the projective norm on $ V_1 \otimes \cdots\otimes V_m$ is denoted $ V_1 \otimes_{\pi} \cdots\otimes_{\pi} V_m$.

\end{definition}
The projective norm of a tensor can also be defined as 
\begin{equation} \label{eq:def-projective-norm-2}
\|\mathcal{T}\|_\pi := \inf \left\{ \sum_{k=1}^{r_N} |\lambda_k| \, : \, r_N \in \mathbb N, \mathcal{T} = \sum_{k=1}^{r_N} \lambda_k\, x_k^1 \otimes \cdots \otimes x_k^m, \ x_k^i\in V_i,\ \|x_k^i\|_{V_i}\leq 1 \right\}.
\end{equation}
 The number $r_N$ is called the nuclear rank, i.e. the number for which it holds such a minimum nuclear decomposition of the tensor $\mathcal{T}$. Given that the quantity $ \sum\limits_{k=1}^{r} \prod\limits_{j=1}^k \|x_k^j\|_{V_i} $ can be seen as the \emph{energy} of the tensor $\mathcal{T}=\sum_{k=1}^{r} x_k^1 \otimes \cdots \otimes x_k^m,$ the projective norm can be interpreted as the \textbf{minimum energy} to decompose the tensor $\mathcal{T}$ into a sum of rank-one tensors \cite{BFZ23}. 
\begin{definition}
   The \emph{injective (spectral) tensor norm} of the tensor $\mathcal{T} \in V_1 \otimes \cdots\otimes V_m$ is
\begin{equation}
\label{eq:def-injective-norm}\|\mathcal{T}\|_\epsilon := \sup \left\{\left| \ip{ \alpha^1 \otimes \cdots \otimes \alpha^k}{\mathcal{T}}\right| \, : \, \alpha^i \in V_i^*,\  \|\alpha^i\|_{V_i^*} \leq 1 \right\},
 \end{equation} 
 Here, $V_i^*$ is the dual space of $V_i,$ that is, the space of all bounded linear functionals on $V_i$, endowed with the norm $\|\alpha^i\|_{V_i^*}:=\sup\limits_{\|x\|_{V_i}\leq 1}|\alpha(x)|$. The Banach space induced by the injective norm on $ V_1 \otimes \cdots\otimes V_m$ is denoted by $V_1 \otimes_{\epsilon} \cdots\otimes_{\epsilon} V_m$.
\end{definition}
The nuclear and spectral norms factorize on simple tensors, i.e.  for all $x_1\in V_1,\ldots,x_m\in V_m$,
$$\|x_1 \otimes \cdots \otimes x_m \|_\pi = \|x_1 \otimes \cdots \otimes x_m \|_\epsilon = \|x_1\|_{V_1} \cdots \|x_m\|_{V_m},$$
and they are  dual to one another, that is for all $\mathcal{T} \in V_1 \otimes \cdots \otimes V_m$, 
\begin{align*}
     \|\mathcal{T}\|_{\pi} &= \sup_{\|\alpha\|_{V_1^* \otimes_\epsilon \cdots \otimes_\epsilon V_k^*} \leq 1} \ip{ \alpha }{\mathcal{T}} ,\\
    \|\mathcal{T}\|_{\epsilon} &= \sup_{\|\alpha\|_{V_1^* \otimes_\pi \cdots \otimes_\pi V_k^*} \leq 1} \ip{ \alpha }{\mathcal{T}}.
\end{align*}
Among the set of tensor norms that can be defined,  the projective and the injective tensor norms are extremal  \cite[page 127]{Ryan02}, i.e. for any other tensor norm $\|\cdot\|$ on $V_1 \otimes \cdots \otimes V_m$, it holds
$$\forall\ \mathcal{T} \in V_1 \otimes \cdots \otimes V_m, \quad \|\mathcal{T}\|_\epsilon \leq \|\mathcal{T}\| \leq \|\mathcal{T}\|_\pi.$$

In the last years, tensor norms have been used intensively in quantum information theory, for studying different notions or properties of quantum systems, such as entanglement, quantum measurements \cite{LN22}. We recall here basic notions, such as quantum entanglement, which will be relevant for the goal of this paper.  

A \emph{pure multipartite quantum state} is a unit vector $\ket{\psi}$ in a tensor product of complex Hilbert spaces $H_1 \otimes \cdots \otimes H_k$, where $k\geq 2$ and $H_i\cong\mathbb C^{d_i}$ for $1\leq i\leq k$. The vector $\ket{\psi}$ is said to be \emph{separable} if it is a product tensor:
$$\ket \psi = \ket{\psi_1} \otimes \cdots \otimes \ket{\psi_k}.$$

Here, each Banach space which appears in the factorization is $(H_i, \|\cdot\|_2)$, where $\|\cdot\|_2 $ is the Euclidean norm. Usually, $\ell_2^d := (\mathbb C^d, \|\cdot\|_2)$. The separability condition for pure states is described in a simple manner, as follows

\begin{proposition} \label{prop:sep-pure}
    A multipartite pure  state $\ket{\psi} \in H_1 \otimes \cdots \otimes H_k$, $\|\psi\|_2=1$, is separable if and only if 
    $$\|\psi\|_\epsilon = \|\psi\|_\pi = 1.$$
\end{proposition}

For mixed multipartite quantum states, which are positive semidefinite and unit trace operators $\rho$ on $H_1\otimes\cdots\otimes H_k$, each of the Banach spaces $\mathcal B(H_i)\cong \mathcal M_{d_i}(\mathbb C)$, $1\leq i\leq k$, is considered with respect to the  \emph{Schatten $1$-norm} (or \emph{nuclear norm})
$$\|X\|_1 = \Tr\sqrt{X^*X}.$$
We write $S_1^{d} := (\mathcal M_d(\mathbb C), \|\cdot\|_1)$ for the complex Banach space. Since mixed quantum states are self-adjoint operators, we shall also consider the real Banach space $S_{1,sa}^d := (\mathcal M_d^{sa}(\mathbb C), \|\cdot\|_1)$.

The separability of mixed quantum states can be checked using the following  
\begin{theorem}\cite{R2000} 
	\label{th:projective-norm}For a multipartite mixed quantum state $\rho \in \mathcal M_{d_1}(\mathbb C) \otimes \cdots \otimes \mathcal M_{d_m}(\mathbb C)$, $\rho \geq 0$, $\Tr \rho =1$, the following assertions are equivalent:
	\begin{enumerate}
	    \item[i)] $\rho$ is separable,
	    \item[ii)] $\|\rho\|_{S_{1,sa}^{d_1} \otimes_\pi \cdots \otimes_\pi S_{1,sa}^{d_m}} = 1$,
	    \item[iii)] $\|\rho\|_{S_1^{d_1} \otimes_\pi \cdots \otimes_\pi S_1^{d_m}} = 1$.
	\end{enumerate}
\end{theorem}
Theorem \eqref{th:projective-norm} gives a necessary and sufficient condition to check separability, but is difficult to use in practice as the computation of tensor norms is an NP-hard problem. 

For pure states $\ket{\psi}\in (\mathbb{C}^d)^{\otimes m}$, it holds that \cite{JLN20}[Remark12.3]
\begin{equation}
  \|\ket{\psi}\|^2_{(\ell_2^d)^{\otimes _{\pi}m}}=\|\kb{\psi}{\psi}\|_{(S_1^{d})^{\otimes_{\pi}m}}
\end{equation}
Also, the injective tensor norm is related to the geometric measure of entanglement \cite{FLN24}, which is a faithful measure of entanglement for multipartite pure states (it is equal to 0 if and only if the state is a product state).

\section{Algorithm}
\label{S:Alg}
In this section, we briefly explain the existing algorithm used to compute the projective norm. We then elaborate upon our developed method, which we will use to estimate the projective norm. Our method is further extended to symmetric tensors and density matrices.

\subsection{Alternating Method}

The alternating method algorithm \cite{DFLW17} computes the projective norm of a tensor by optimizing over a single rank-one product state tensor at a time while keeping the rest of the rank-one product state tensors fixed. This results in a minimization problem that is equivalent to minimizing the sum of the Euclidean norms under a linear constraint—specifically a {\em second order cone programming} (SOCP). For symmetric tensors, the algorithm is adapted to maintain symmetry throughout the decomposition, further utilizing the symmetric nature of tensor structure to reduce computational overheads.

This method provides an upper bound on the projective norm. However, due to the alternating approach, it does not guarantee a global optimum, and might even converge to a local minimum or critical point.

\subsection{Canonical Polyadic Decomposition}

Before we explain our algorithm, it is crucial to understand the Canonical Polyadic Decomposition (CPD) \cite{KB09}, as the next sections expand on this to build up our algorithm. The CPD attempts to approximate a tensor $\mathcal{T}$ as a sum of $R$ rank-one tensors, $R$ being a finite positive integer. The vectors $a_j^i$ $\forall i \in \{1, 2,...,m\}$ involved in calculating the rank one tensors are called cores.

\begin{equation}
    \label{eq:CPD-definition}
    \mathcal{T}' = \sum_{j=1}^R a_j^1\otimes a_j^2\otimes\cdots a_j^m
\end{equation}

The goal of CPD is to minimize the mean square error $\|\mathcal{T - T'}\|$ (we shall call this the {\em reconstruction cost} henceforth) by optimizing the cores. One such algorithm to achieve this is the {\em alternating least squares} (ALS) \cite{Har70} algorithm where the optimization is performed between specific views of the approximation and the original tensor. However, the rate of convergence is slow and might not also converge to a global optimum owing to the fact that the ALS considers only one view of the original tensor at a time. Hence we use a stochastic gradient decent algorithm which adjusts parameters by computing the individual adaptive step sizes for different parameter estimates of the first and second order gradients \cite{KB17}. This allows simultaneous optimization of all the cores and overcoming saddle points on the parameter landscape by adaptively adjusting the step size such that the algorithm is not stuck in a shallow local minima. This also improves the speed of convergence, especially in higher-dimensional parameter spaces.

\subsection{Adaptive Rank Canonical Polyadic Decomposition}

The CPD approximates a tensor $\mathcal{T}$ as a sum of $R$ rank-one tensors. This does not give a minimum rank decomposition. To ensure that $R$ converges to the rank of a tensor $r(\mathcal{T})$, we make modifications to the objective function of CPD to optimize not only the cores but also $R$, which we call as the {\em adaptive rank canonical polyadic decomposition} (ARCPD). We do this by introducing coefficients $C_j$ to our reconstructed tensor $\mathcal{T}'$.

\begin{equation}
    \label{eq:ARCPD-initialisation}
    \mathcal{T}' = \sum_{j=1}^R C_j\cdot \phi_j,\text{\hspace{0.5cm}where }\phi_j = \frac{a_j^1\otimes a_j^2\otimes\cdots a_j^m}{\|a_j^1\otimes a_j^2\otimes\cdots a_j^m\|}
\end{equation}

We optimize such that we minimize not just the reconstruction cost but also the number of $C_j$ terms going to zero (we shall call this the {\em rank cost} henceforth). Hence the effective rank is the number of $C_j$ coefficients greater than a small threshold. We call this algorithm {\em adaptive} because we start with a decomposition of a higher rank $R$ and adaptively reduce to the minimal rank $r(\mathcal{T})$. We chose $R$ as the strong upper bound for the rank of a tensor \cite{BFZ23} given by

\begin{equation*}
    r(\mathcal{T}) \le \frac{\prod_{i=1}^m d_i}{\text{max}\{d_1, d_2,...,d_m\}} = R
\end{equation*}
where $d_i$ is the dimensionality of the Banach space $V_i$.

The objective function $\mathcal{L}_R(\mathcal{T}, a, C)$ to minimize for this decomposition is given as

\begin{equation}
    \label{eq:ARCPD-objective-fn}
    \mathcal{L}_R(\mathcal{T}, a, C) = k_1\|\mathcal{T - T'}\| + k_2\sum_{j=1}^R 1_{\{|C_j| > \epsilon\}}
\end{equation}
where $\epsilon \rightarrow0$ is the small threshold that forces the maximum number of coefficients to become as close to zero, and $k_1, k_2$ are {\em regularization constants}. As we desire the reconstruction cost and rank cost to be strongly minimized, we chose large values of $k_1$ and $k_2$ to achieve this optimization.

\subsection{Nuclear Rank Canonical Polyadic Decomposition}

The nuclear rank is the minimum rank of the tensor decomposition which gives us the projective norm as seen from Eq. \ref{eq:def-projective-norm} and Eq. \ref{eq:def-projective-norm-2}. If we decompose our tensor $\mathcal{T}$ for the minimum rank and the sum of norms of the corresponding product state coefficients (we shall call this the {\em norm cost}) simultaneously, we will converge to the nuclear rank and the norm cost will converge to the projective norm. This comes from the following proposition \ref{prop:nuclear-rank} and its corollary \ref{corollary:nuclear-rank}.

\begin{proposition}
    \label{prop:nuclear-rank} \cite{FL14} Let V be a real vector space of dimension n and $\nu: V \rightarrow [0, \infty)$ be a norm. Suppose $\mathcal{E}$, the set of the extreme points of the unit ball $B_\nu$, is compact. Then the nuclear rank $\textup{rank}_\nu(x_m): V \rightarrow \mathbb{R}$ is an upper semi-continuous function, i.e., if $(x_m)^\infty_{m=1}$ is a convergent sequence in $V$ with $\textup{rank}_\nu(x_m)\le r$ for all $m \in \mathbb{N}$, then $x = \textup{lim}_{m\rightarrow\infty}x_m$ must have $\textup{rank}_\nu \le r$.
\end{proposition}

\begin{corollary}
    \label{corollary:nuclear-rank} For any $A \in \mathbb{R}^{n_1\times \cdots\times n_m}$, the best nuclear rank-r approximation problem
    \begin{equation*}
        \label{eq:nuclear-rank-approximation-problem}
        \textup{argmin}\{\|A - X\|: \textup{rank}_*(X)\le r\}
    \end{equation*}
    always has a solution.
\end{corollary}

Incorporating the norm cost into the objective function of the ARCPD, we obtain the {\em nuclear rank canonical polyadic decomposition} (NRCPD). The objective function $\mathcal{L}_N(\mathcal{T}, a, C)$ for this decomposition is given as

\begin{equation}
    \label{eq:NRCPD-objective-fn}
    \mathcal{L}_N(\mathcal{T}, a, C) = k_1\|\mathcal{T - T'}\| + k_2\sum_{j=1}^R 1_{\{|C_j| > \epsilon\}} + k_3\sum_{j=1}^R |C_j|
\end{equation}

\begin{algorithm}
    \label{alg:NRCPD}
    \caption{Nuclear Rank Canonical Polyadic Decomposition to estimate projective norm of $\mathcal{T}$}
    \begin{algorithmic}[1]
        \Require $\|\mathcal{T - T'}\| \rightarrow 0$
        \Ensure $a_j^i \ne 0$ for $\mathcal{T}\ne0$
        \State $R \gets \frac{\prod_{i=1}^m d_i}{\text{max}\{d_1, d_2,...,d_m\}}$
        \State \textbf{Intialize:} $a_j^i \sim \mathcal{N}\left(0, \frac{1}{d_i}\right), \,\forall i \in 1, 2,\cdots,m, \,\forall j\in 1, 2, \cdots, R$
        \State \textbf{Initialize:} $C_j \sim \mathcal{N}(0, 1),\,\forall j \in 1, 2, \cdots R$
        \If{$\mathcal{T}$ is complex}
            \State $\tilde{a}_j^i \sim \mathcal{N}\left(0, \frac{1}{d_i}\right), \,\forall i \in 1, 2,\cdots,m, \,\forall j\in 1, 2, \cdots, R$
            \State $\tilde{C}_j \sim \mathcal{N}(0, 1),\,\forall j \in 1, 2, \cdots R$
            \State $a_j^i \gets a_j^i + i.\tilde{a}_j^i,\,\forall i \in 1, 2,\cdots,m, \,\forall j\in 1, 2, \cdots, R$
            \State $C_j \gets C_j + i.\tilde{C}_j,\,\forall j\in 1, 2, \cdots, R$
        \EndIf
        \State $t = 0$
        \For{$t<$\texttt{max epochs}}
            \State $\phi_j \gets \frac{a_j^1\otimes a_j^2\otimes\cdots a_j^m}{\|a_j^1\otimes a_j^2\otimes\cdots a_j^m\|},\,\forall j\in 1,2,\cdots,R$
            \State $\mathcal{T}' \gets \sum_{j=1}^R C_j\cdot\phi_j$
            \State $\mathcal{J} \gets \mathcal{L}_N(\mathcal{T}, a, C)$
            \State $a_j^i \gets a_j^i - \alpha\frac{\partial\mathcal{J}}{\partial a_j^i}, \,\forall i \in 1, 2,\cdots,m, \,\forall j\in 1, 2, \cdots, R$ \Comment{$\alpha$ is the step size}
            \State $C_j \gets C_j - \alpha\frac{\partial\mathcal{J}}{\partial C_j}, \,\forall j\in 1, 2, \cdots, R$
            \State $t\gets t+1$
        \EndFor
        \State $r_N = 0$
        \If{$C_j >$ \texttt{tolerance}}
            \State $r_N \gets r_N + 1, \, \forall j \in 1, 2, \cdots,R$
        \EndIf
        \State $\|\mathcal{T}\|_\pi = \sum_{j=1}^R |C_j|$
    \end{algorithmic}
\end{algorithm}

\newpage

\subsection{Symmetric Nuclear Rank Canonical Polyadic Decomposition}

A tensor $\mathcal{T}^{(s)} \in V_1\otimes\cdots\otimes V_m$ is said to be symmetric if $d_1=\cdots=d_m = d_s$ and $\mathcal{T}^{(s)}_{i_1\cdots i_m} = \mathcal{T}^{(s)}_{j_1\cdots j_m}$ where $(i_1, \cdots, i_m)$ is a permutation of $(j_1, \cdots, j_m)$ \cite{Nie17}. For every $m$-partite symmetric tensor $\mathcal{T}^{(s)}$, there exists a decomposition \cite{CGLM08} of the form

\begin{equation}
    \label{eq:symmetric-tensor-decomposition}
    \mathcal{T}^{(s)} = \sum_{j=1}^{r_S} \otimes_{i=1}^m x_j
\end{equation}

where $r_S$ is the {\em symmetric rank}. Symmetric tensors find applications in various fields such as in machine learning, communications, signal, speech and image processing \cite{Fav21}. In physics, symmetric quantum states are also known as {\em bosons} \cite{FK16}.

From Eq. \ref{eq:symmetric-tensor-decomposition}, it is clear that due to the symmetric nature of the tensor, it has a reduced parameter space which simplifies the calculations and computations, providing tighter bounds for finding the symmetric rank, nuclear rank and projective norm. Due to
being less entangled, their geometric measure of entanglement is polynomially computable. This simplifies the NRCPD to obtain the {\em symmetric nuclear rank canonical polyadic decomposition} (SNRCPD). Therefore the objective function $\mathcal{L}^{(s)}_N$ for computing the projective norm of a symmetric tensor simplifies as

\begin{equation}
    \label{eq:SNRCPD-objective-fn}
    \mathcal{L}^{(s)}_N\left(\mathcal{T}^{(s)}, a, C\right) = k_1\|\mathcal{T}^{(s)} - \mathcal{T'}\| + k_2\sum_{j=1}^R 1_{\{|C_j| > \epsilon\}} + k_3\sum_{j=1}^R |C_j|
\end{equation}

\begin{algorithm}
    \label{alg:SNRCPD}
    \caption{Symmetric Nuclear Rank Canonical Polyadic Decomposition to estimate projective norm of $\mathcal{T}^{(s)}$}
    \begin{algorithmic}[1]
        \Require $\|\mathcal{T}^{(s)} - \mathcal{T'}\| \rightarrow 0$
        \Ensure $a_j \ne 0$
        \State $R \gets d_s^{m-1}$
        \State \textbf{Intialize:} $a_j \sim \mathcal{N}\left(0, \frac{1}{d_s}\right), \,\forall j\in 1, 2, \cdots, R$
        \State \textbf{Initialize:} $C_j \sim \mathcal{N}(0, 1),\,\forall j \in 1, 2, \cdots R$
        \If{$\mathcal{T}$ is complex}
            \State $\tilde{a}_j \sim \mathcal{N}\left(0, \frac{1}{d_s}\right), \,\forall j\in 1, 2, \cdots, R$
            \State $\tilde{C}_j \sim \mathcal{N}(0, 1),\,\forall j \in 1, 2, \cdots R$
            \State $a_j \gets a_j + i.\tilde{a}_j,\,\forall j\in 1, 2, \cdots, R$
            \State $C_j \gets C_j + i.\tilde{C}_j,\,\forall j\in 1, 2, \cdots, R$
        \EndIf
        \State $t = 0$
        \For{$t<$\texttt{max epochs}}
            \State $\phi_j^{(s)} \gets \frac{\otimes_{i=1}^m a_j}{\|\otimes_{i=1}^m a_j\|},\forall j\in 1,2,\cdots,R$
            \State $\mathcal{T}' \gets \sum_{j=1}^R C_j\cdot \phi_j^{(s)}$
            \State $\mathcal{J} \gets \mathcal{L}_N(\mathcal{T}^{(s)}, a, C)$
            \State $a_j \gets a_j - \alpha\frac{\partial\mathcal{J}}{\partial a_j}, \,\forall j\in 1, 2, \cdots, R$ \Comment{$\alpha$ is the step size}
            \State $C_j \gets C_j - \alpha\frac{\partial\mathcal{J}}{\partial C_j}, \,\forall j\in 1, 2, \cdots, R$
            \State $t\gets t+1$
        \EndFor
        \State $r_N = 0$
        \If{$C_j >$ \texttt{tolerance}}
            \State $r_N \gets r_N + 1, \, \forall j \in 1, 2, \cdots,R$
        \EndIf
        \State $\|\mathcal{T}^{(s)}\|_\pi = \sum_{j=1}^R |C_j|$
    \end{algorithmic}
\end{algorithm}

\subsection{Nuclear Rank Canonical Polyadic Density Matrix Decomposition}

Alongside tensors, quantum states can also be represented using the density matrix formulation. We shall first define the representation and notation we shall be using before defining the projective norm for a density matrix.

\begin{definition}
    \cite{NC10} Given a $m$-partite state $|\psi_k\rangle \in H_1 \otimes\cdots\otimes H_m$, where $k$ is the index with respective probabilities $p_k$ as part of the ensemble of pure states $\{p_k,\,|\psi_k\rangle\}$, the density matrix for the system is defined as
    \begin{equation}
        \label{eq:density-matrix-def}
        \rho \equiv \sum_k p_k|\psi_k\rangle\langle\psi_k|,\,\text{such that }\sum_k p_k = 1
    \end{equation}
\end{definition}

In such a representation, a state $\rho$ is said to be {\em separable} if it can be written as a convex combination of the product density matrices as follows

\begin{equation*}
    \label{eq:separable-density-matrix-def}
    \rho = \sum_{k = 1}^r \lambda_k\cdot\rho_1^{(k)}\otimes\cdots\otimes\rho_m^{(k)}
\end{equation*}

for a probability distribution $(\lambda_i)_{k=1}^r$ and $\rho_j^{(k)} \in S_1^{d_j}$ are normalized. Otherwise $\rho$ is said to be {\em entangled}. With this, we define the projective norm for a density matrix as

\begin{equation}
    \label{eq:density-matrix-projective-norm-def}
    \|\rho\|_\pi = \text{inf}\left\{\sum_{k=1}^{r_N^{(D)}}|\lambda_k|\,:\, r_N^{(D)} \in \mathbb N,\,\rho = \sum_{k=1}^{r_N^{(D)}}\lambda_k\cdot\rho_1^{(k)}\otimes\cdots\otimes\rho_m^{(k)},\,\rho_j^{(k)}\in S_1^{d_j},\, \|\rho_j^{(k)}\|_1 =1 \right\}
\end{equation}

where $r_N^{(D)}$ if the {\em nuclear rank of the density matrix}.

We extend the NRCPD to handle density matrices by increasing the parameter space by optimizing over the coefficients $C_j$, the kets $a_j^{(i)}$, and the bras $b_j^{(i)}$ (which are not normalized), which allows our decomposition to optimize over all possible combinations of the product state density matrices.

\begin{equation}
    \label{eq:NRCPDMD-def}
    \rho' = \sum_{j=1}^{R^{(D)}} C_j\cdot \Phi_j,\text{\hspace{0.5cm}where }\Phi_j = \frac{\otimes_{i=1}^{m}|a_j^i\rangle\langle b_j^i|}{\|\otimes_{i=1}^{m}|a_j^i\rangle\langle b_j^i|\|_1}
\end{equation}

where $R^{(D)}$ is the upper bound for the rank of a density matrix which increases owing to the increased parameter space from including both the kets and bras.

\begin{equation*}
    R^{(D)} = \left(\frac{\prod_{i=1}^m d_i}{\text{max}\{d_1, d_2,...,d_m\}}\right)^2
\end{equation*}

The objective function $\mathcal{L}_N^{(D)}$ to compute the projective norm of a density matrix is given as

\begin{equation}
    \label{eq:NRCPDMD-objective-fn}
    \mathcal{L}_N^{(D)}(\rho, a, b, C) = k_1\|\rho - \rho'\| + k_2\sum_{j=1}^{R^{(D)}} 1_{\{|C_j| > \epsilon\}} + 
    k_3 \sum_{j=1}^{R^{(D)}} |C_j|
\end{equation}

\begin{algorithm}
    \label{alg:NRCPDMD}
    \caption{Nuclear Rank Canonical Polyadic Density Matrix Decomposition to estimate projective norm of $\rho$}
    \begin{algorithmic}[1]
        \Require $\|\rho - \rho'\| \rightarrow 0$
        \Ensure $a_j^i,\, b_j^i \ne 0$
        \State $R^{(D)} \gets \left(\frac{\prod_{i=1}^m d_i}{\text{max}\{d_1, d_2,...,d_m\}}\right)^2$
        \State \textbf{Intialize:} $a_j^i \sim \mathcal{N}\left(0, \frac{1}{d_i}\right), \,\forall i \in 1, 2,\cdots,m, \,\forall j\in 1, 2, \cdots, R^{(D)}$
        \State \textbf{Intialize:} $b_j^i \sim \mathcal{N}\left(0, \frac{1}{d_i}\right), \,\forall i \in 1, 2,\cdots,m, \,\forall j\in 1, 2, \cdots, R^{(D)}$
        \State \textbf{Initialize:} $C_j \sim \mathcal{N}(0, 1),\,\forall j \in 1, 2, \cdots R^{(D)}$
        \If{$\rho$ is complex}
            \State $\tilde{a}_j^i \sim \mathcal{N}\left(0, \frac{1}{d_i}\right), \,\forall i \in 1, 2,\cdots,m, \,\forall j\in 1, 2, \cdots, R^{(D)}$
            \State $\tilde{b}_j^i \sim \mathcal{N}\left(0, \frac{1}{d_i}\right), \,\forall i \in 1, 2,\cdots,m, \,\forall j\in 1, 2, \cdots, R^{(D)}$
            \State $\tilde{C}_j \sim \mathcal{N}(0, 1),\,\forall j \in 1, 2, \cdots R^{(D)}$
            \State $a_j^i \gets a_j^i + i.\tilde{a}_j^i,\,\forall i \in 1, 2,\cdots,m, \,\forall j\in 1, 2, \cdots, R^{(D)}$
            \State $b_j^i \gets b_j^i + i.\tilde{b}_j^i,\,\forall i \in 1, 2,\cdots,m, \,\forall j\in 1, 2, \cdots, R^{(D)}$
            \State $C_j \gets C_j + i.\tilde{C}_j,\,\forall j\in 1, 2, \cdots, R^{(D)}$
        \EndIf
        \State $t = 0$
        \For{$t<$\texttt{max epochs}}
            \State $\Phi_j = \frac{\otimes_{i=1}^{m}|a_j^i\rangle\langle b_j^i|}{\|\otimes_{i=1}^{m}|a_j^i\rangle\langle b_j^i|\|_1},\forall j\in 1,2,\cdots,R^{(D)}$
            \State $\rho' \gets \sum_{j=1}^{R^{(D)}} C_j\cdot \Phi_j$
            \State $\mathcal{J} \gets \mathcal{L}_N^{(D)}(\rho, a, b, C)$
            \State $a_j^i \gets a_j^i - \alpha\frac{\partial\mathcal{J}}{\partial a_j^i}, \,\forall i \in 1, 2,\cdots,m, \,\forall j\in 1, 2, \cdots, R^{(D)}$ \Comment{$\alpha$ is the step size}
            \State $b_j^i \gets b_j^i - \alpha\frac{\partial\mathcal{J}}{\partial b_j^i}, \,\forall i \in 1, 2,\cdots,m, \,\forall j\in 1, 2, \cdots, R^{(D)}$
            \State $C_j \gets C_j - \alpha\frac{\partial\mathcal{J}}{\partial C_j}, \,\forall j\in 1, 2, \cdots, R^{(D)}$
            \State $t\gets t+1$
        \EndFor
        \State $r_N^{(D)} = 0$
        \If{$C_j >$ \texttt{tolerance}}
            \State $r_N^{(D)} \gets r_N^{(D)} + 1, \, \forall j \in 1, 2, \cdots,R^{(D)}$
        \EndIf
        \State $\|\rho\|_\pi = \sum_{j=1}^{R^{(D)}} |C_j|$
    \end{algorithmic}
\end{algorithm}

\newpage

\section{Computation and Results}
\label{S:computation}

In this section, we showcase the performance of our developed method to calculate the projective norm. We are going to focus on tensors having orders of 2, 3, 4, 5 and 6. The projective norm for order-2 tensors is precisely the Schatten 1-norm of the corresponding matrix, i.e. the sum of the non-zero singular values of the operator. Computing the projective norm for order 3 and higher is an NP-hard problem for which we shall verify our results using the analytical results obtained in \cite{BFZ23}, \cite{FL14} for order 3 tensors and numerical results obtained in \cite{DFLW17} for even higher order tensors. We shall also do the same for density matrices for both pure and mixed states and verify the separability criterion for some given mixed states in \cite{DPS04}. The codes were run on Python using the PyTorch library, which is an open-source machine learning framework that provides efficient tensor operations with GPU acceleration and automatic differentiation, offering dynamic computation graphs and easy manipulation
of higher-order multi-dimensional tensors. PyTorch’s Autograd engine enables automatic gradient computation, using which we can implement various gradient descent-based optimizers such as
Stochastic Gradient Descent (SGD), Adam, RMSProp, etc. \cite{PGML+19}.

\subsection{Projective Norm Computations for Tensors}

\begin{figure}[ht]
    \centering
    \includegraphics[width=\linewidth]{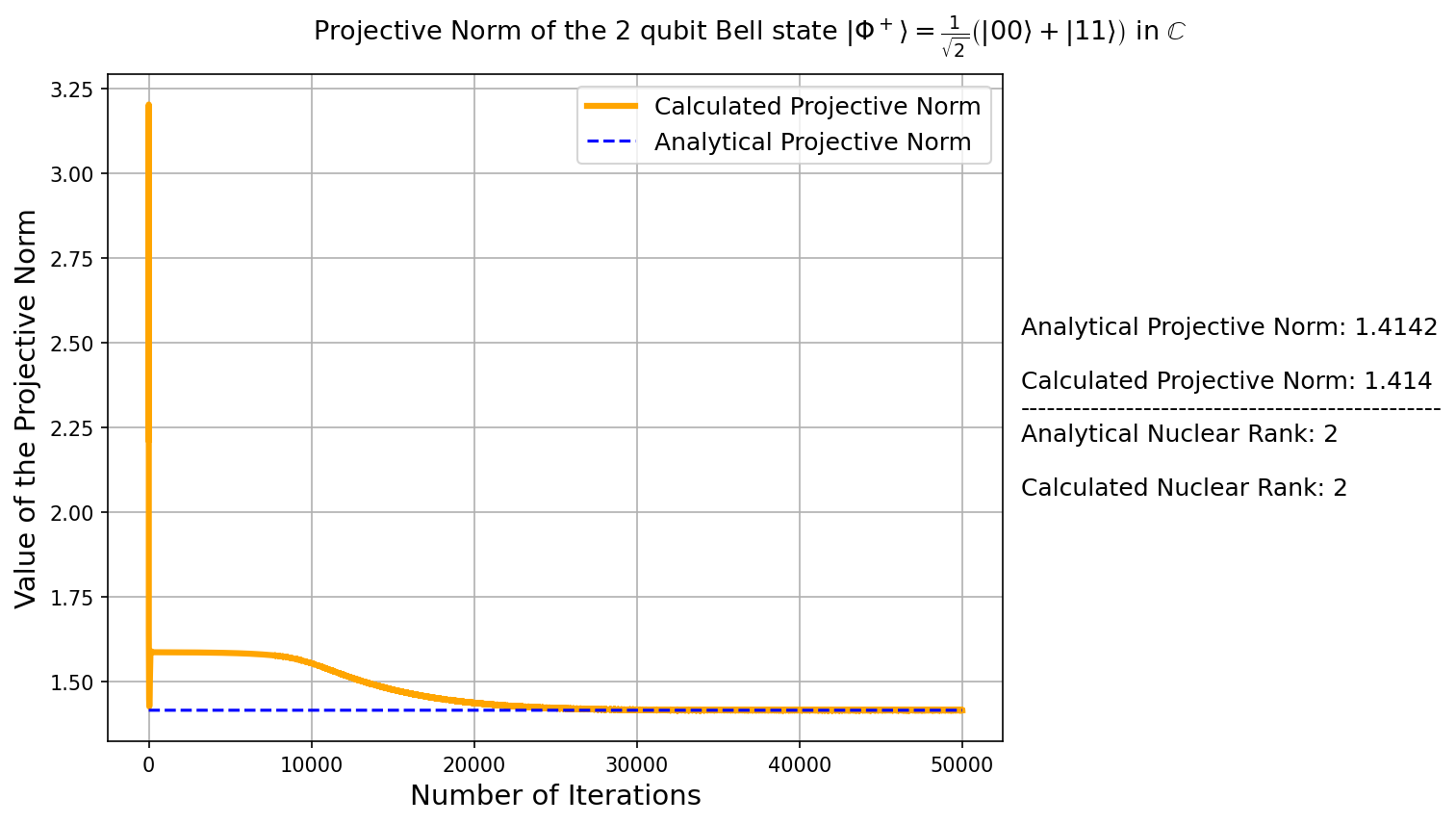}
    \caption{Projective Norm and Nuclear rank of the Bell state $|\phi^+\rangle$ in $\mathbb C$}
    \label{fig:2-qubit-bell-state}
\end{figure}

\begin{figure}[ht]
    \centering
    \includegraphics[width=\linewidth]{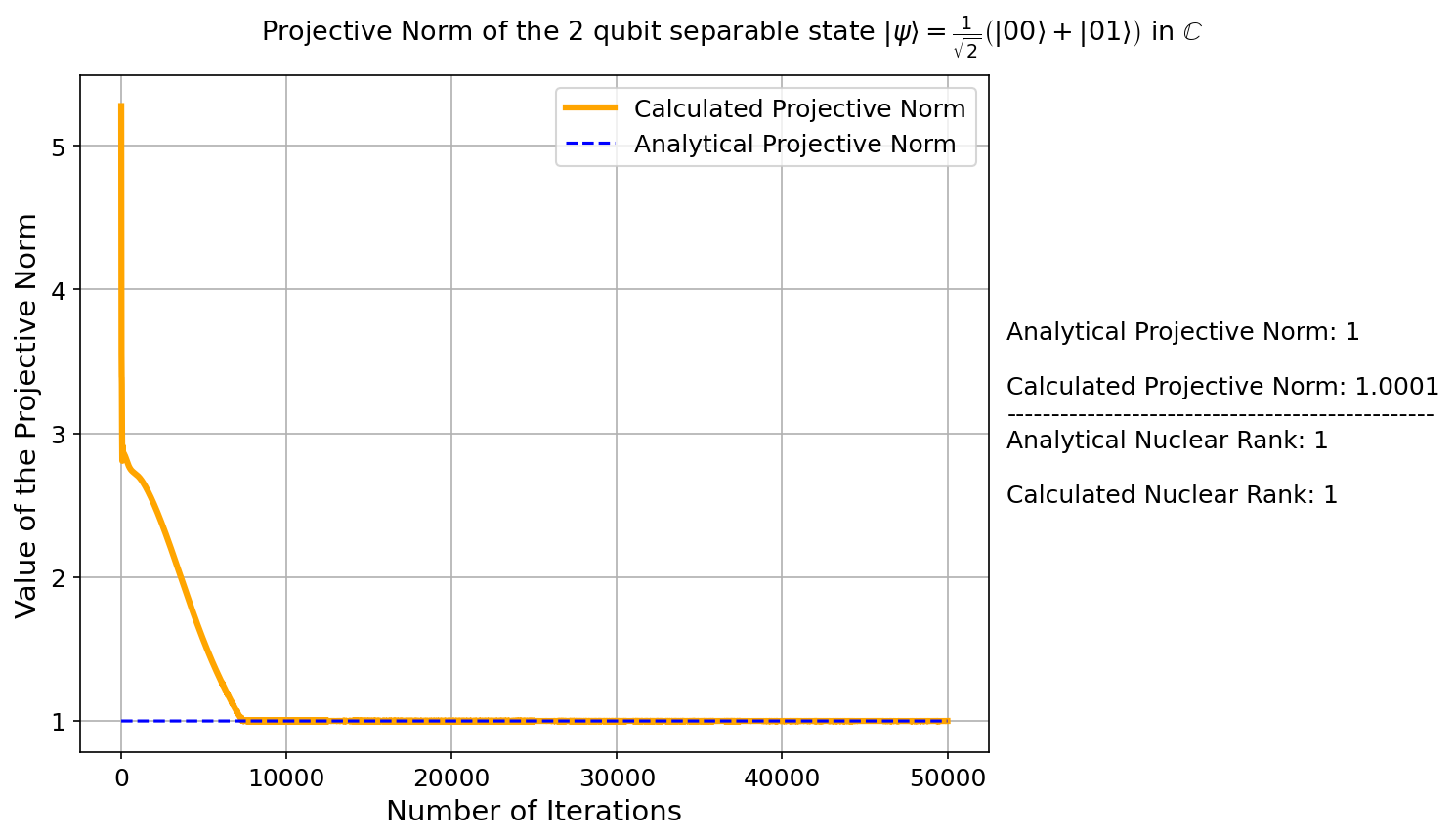}
    \caption{Projective Norm and Nuclear rank of a 2 qubit separable state in $\mathbb C$}
    \label{fig:2-qubit-separable-state}
\end{figure}

The projective norm of 2nd order tensor states (or {\em bipartite states}) can be computed analytically as the sum of the singular values of their respective matrix form. We compute the projective norm and nuclear rank of an entangled state and an arbitrary separable state and compare the numerical results obtained using our algorithm with the analytical values in Fig.\ref{fig:2-qubit-bell-state} and Fig.\ref{fig:2-qubit-separable-state} respectively.

\begin{figure}[ht]
    \centering
    \includegraphics[width = \linewidth]{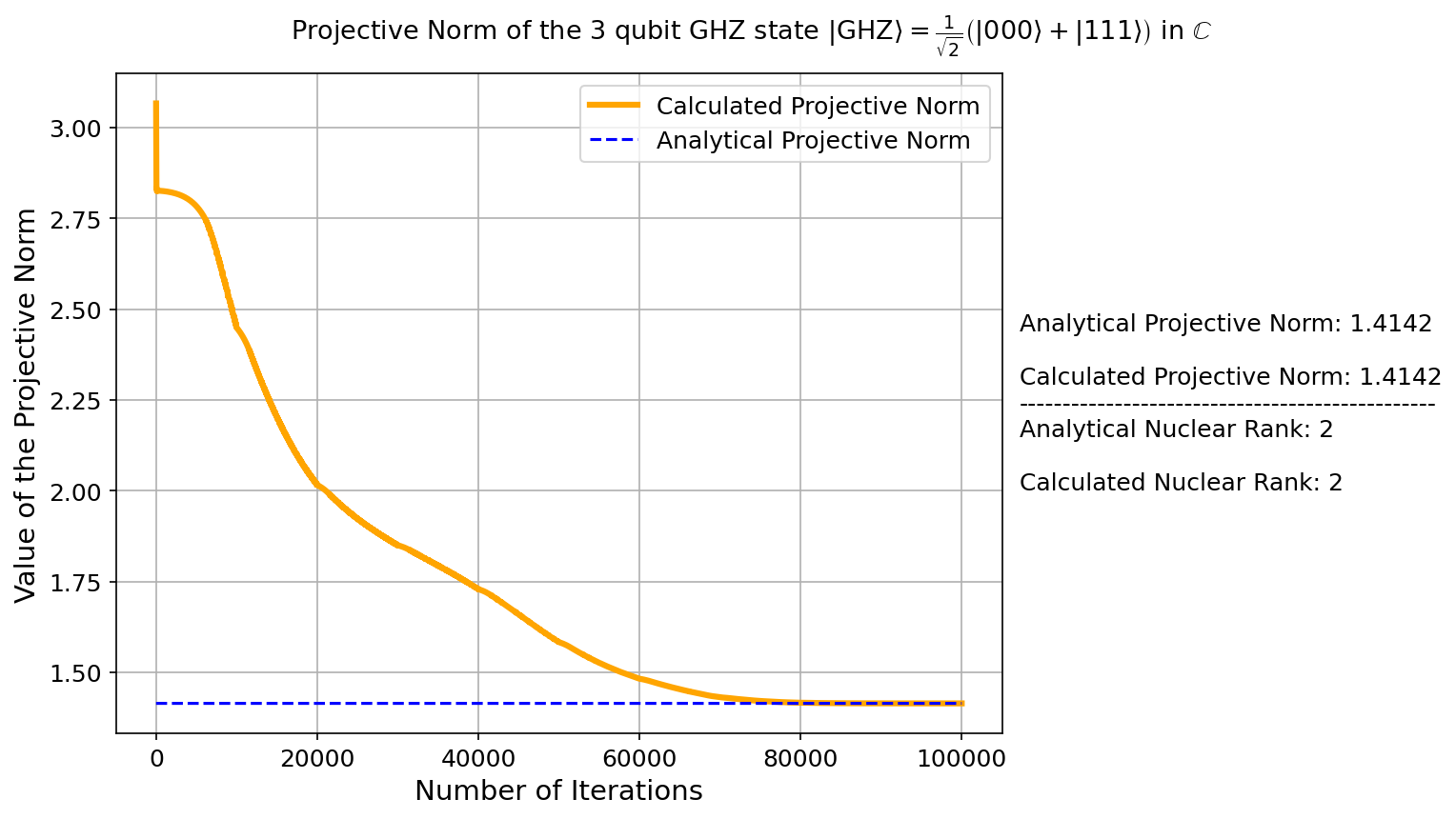}
    \caption{Projective Norm and Nuclear rank of the 3 qubit GHZ state in $\mathbb C$}
    \label{fig:3-qubit-GHZ-state}
\end{figure}

Calculating the projective norm for 3rd order tensor states becomes complicated; however, we have some analytical values of the projective norm calculated for some symmetric states in \cite{BFZ23} and \cite{FL14}. Fig.\ref{fig:3-qubit-GHZ-state} shows the convergence of our numerical results to the analytically calculated projective norm and nuclear rank for the 3-qubit GHZ state.

\newpage

\begin{figure}[ht]
    \centering
    \includegraphics[width=0.9\linewidth]{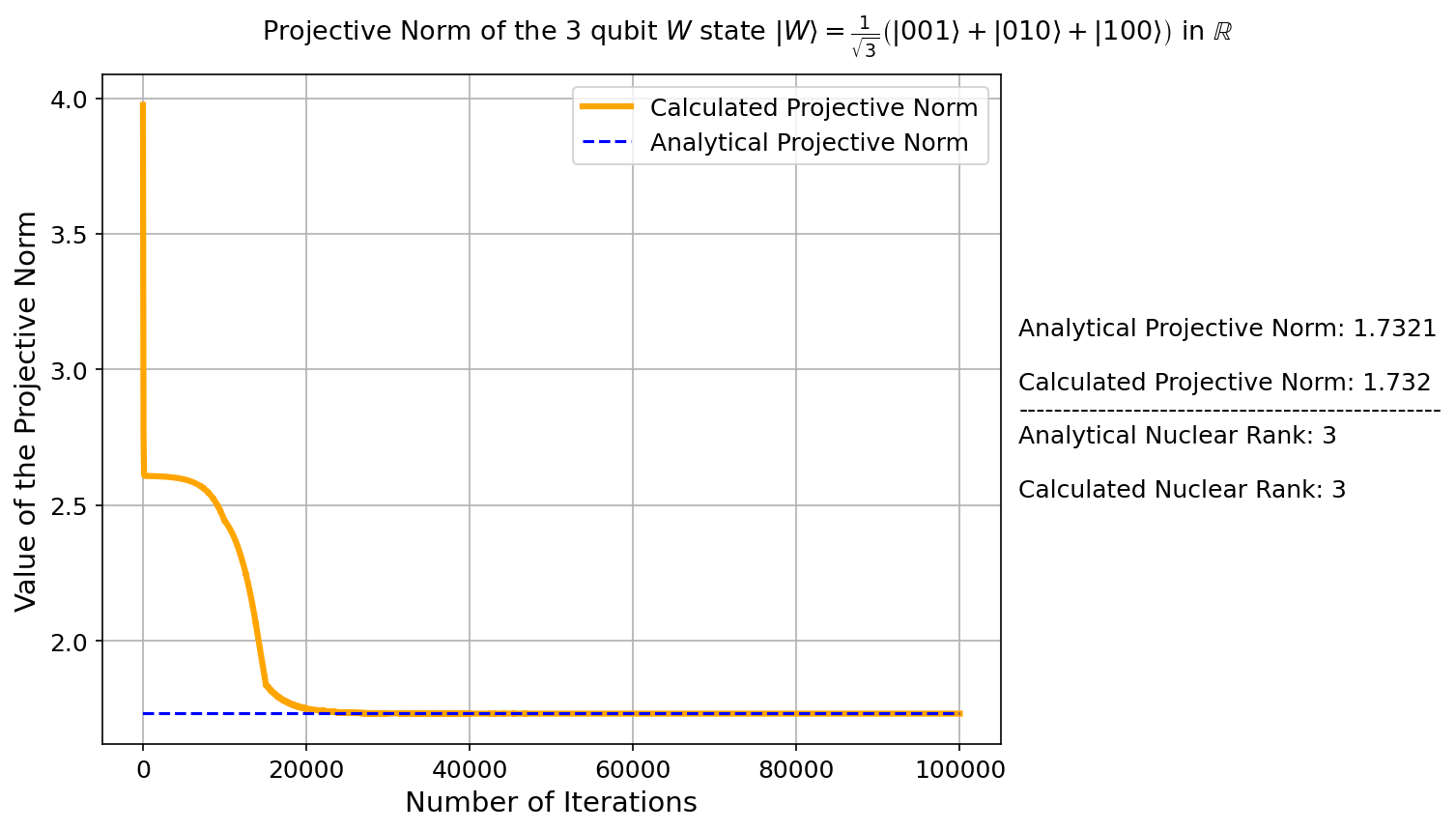}
    \caption{Projective Norm and Nuclear rank of the 3 qubit W state in $\mathbb R$}
    \label{fig:3-qubit-W-state-real}
\end{figure}

\begin{figure}[ht]
    \centering
    \includegraphics[width=0.9\linewidth]{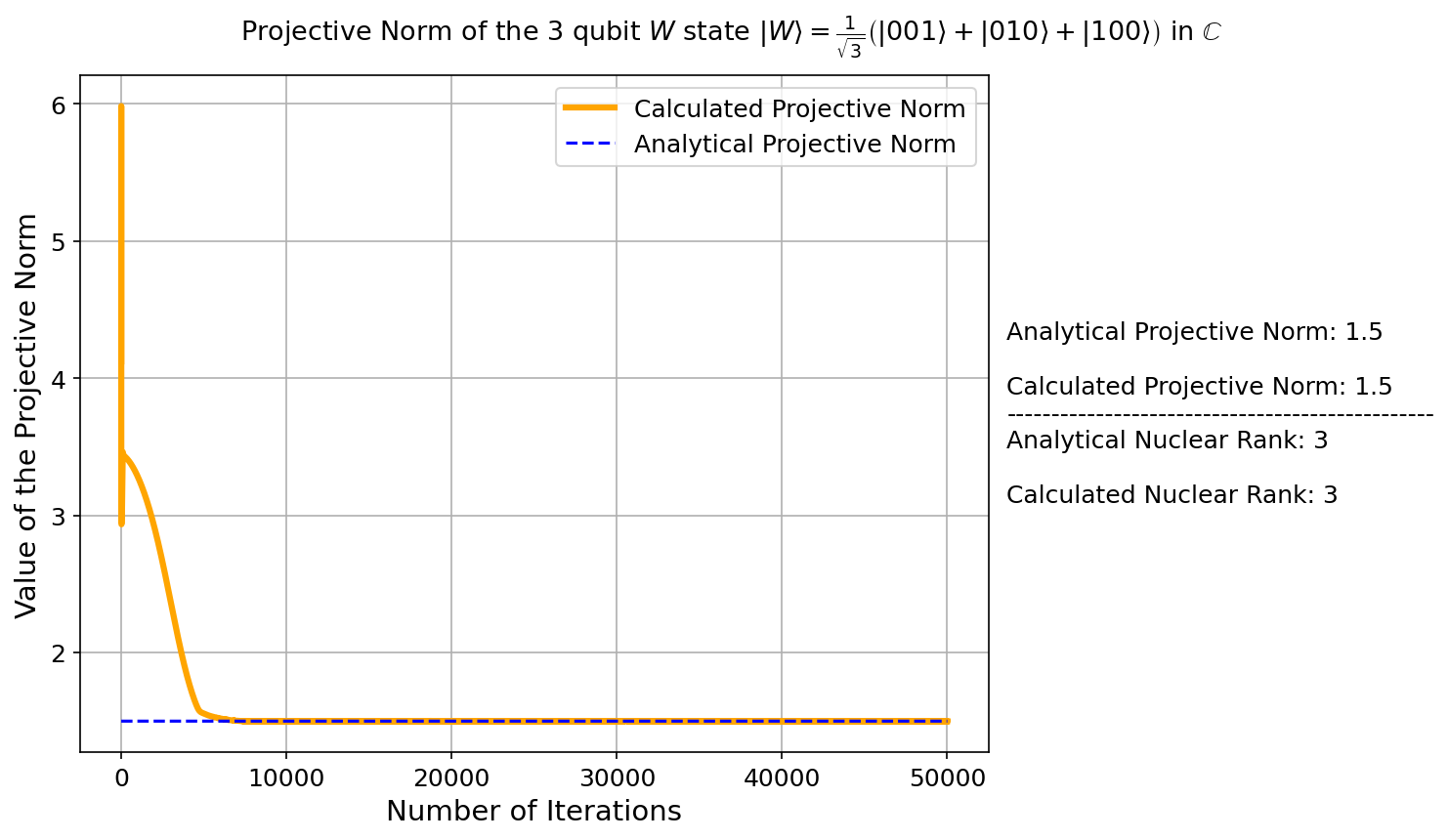}
    \caption{Projective Norm and Nuclear rank of the 3 qubit W state in $\mathbb C$}
    \label{fig:3-qubit-W-state-complex}
\end{figure}

\begin{figure}[ht]
    \centering
    \includegraphics[width= 0.9\linewidth]{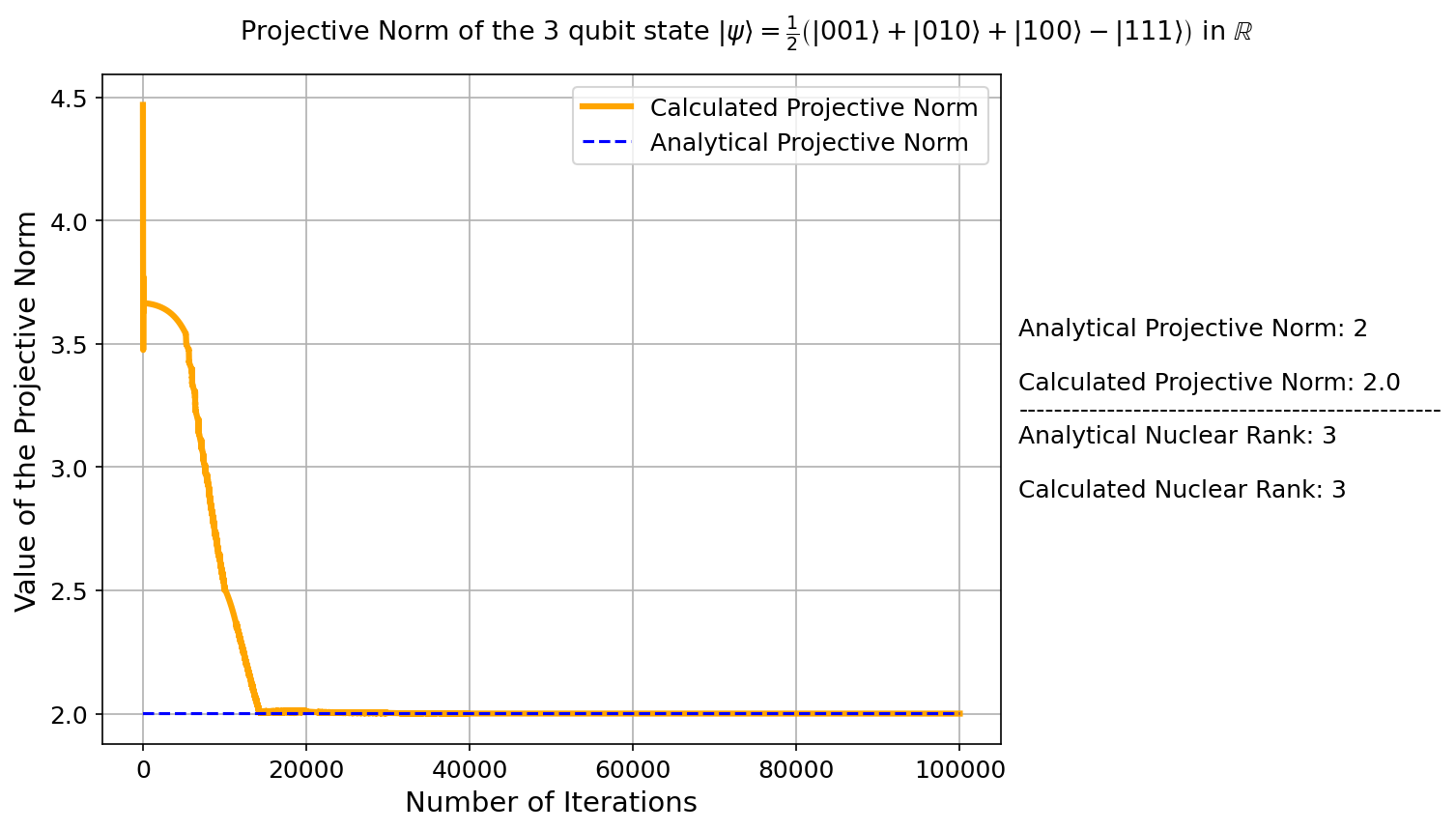}
    \caption{Projective Norm and Nuclear rank of the 3 qubit state in $\mathbb R$}
    \label{fig:3-qubit-B-state-real}
\end{figure}

\begin{figure}[ht]
    \centering
    \includegraphics[width=0.9\linewidth]{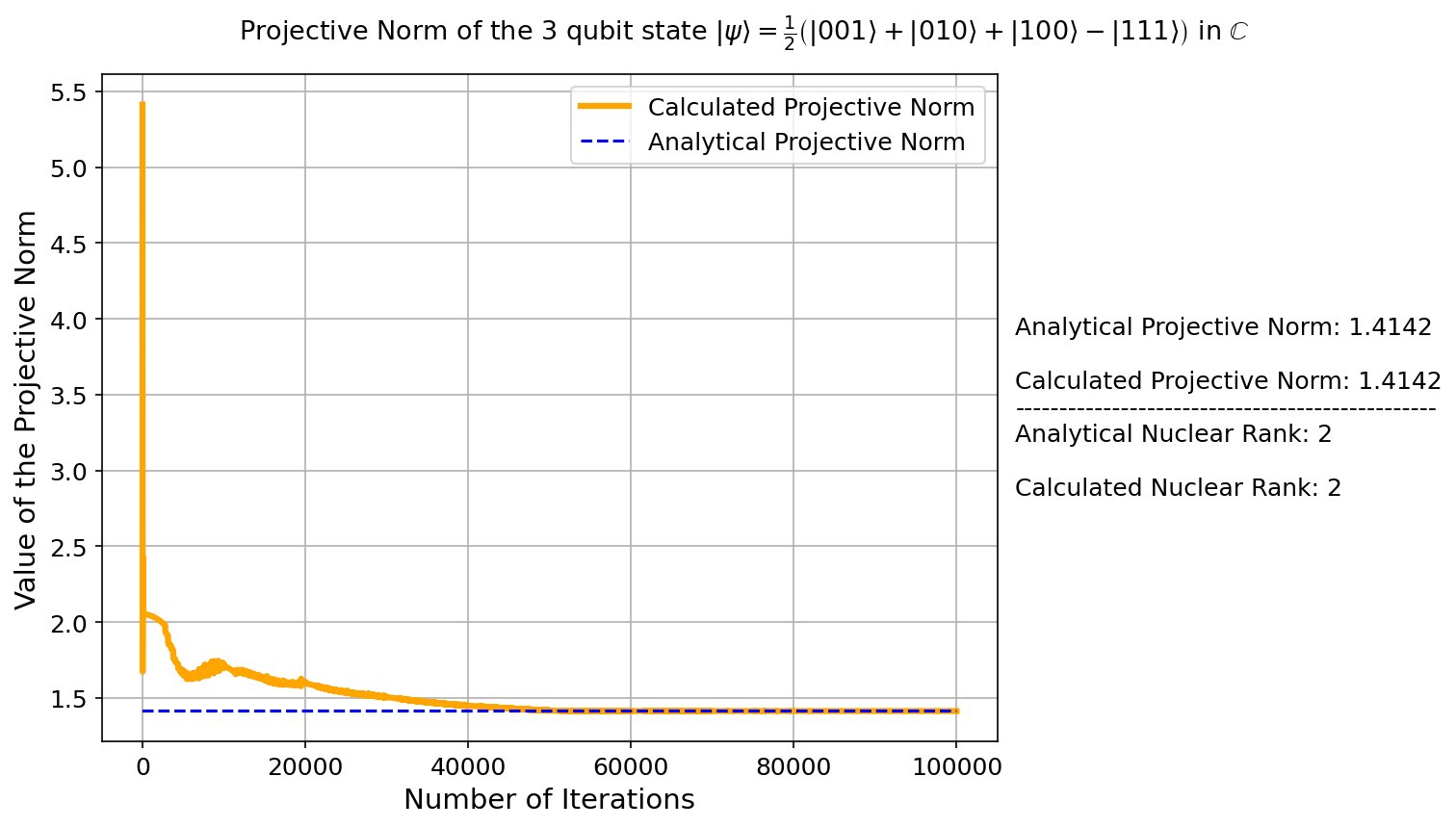}
    \caption{Projective Norm and Nuclear rank of the 3 qubit state in $\mathbb C$}
    \label{fig:3-qubit-B-state-complex}
\end{figure}

In general, a tensor has different projective norms and nuclear ranks over the real and complex fields. We demonstrate this using the 3-qubit W state, and the 3-qubit state $|\psi\rangle = \frac{1}{2}\left(|001\rangle + |010\rangle + |100\rangle - |111\rangle\right)$ using Fig.\ref{fig:3-qubit-W-state-real} and Fig.\ref{fig:3-qubit-W-state-complex} for the W state and Fig.\ref{fig:3-qubit-B-state-real} and Fig.\ref{fig:3-qubit-B-state-complex} for the above mentioned state $|\psi\rangle$ over the real and complex number fields respectively. 

\newpage

\begin{figure}[ht]
    \centering
    \includegraphics[width=0.9\linewidth]{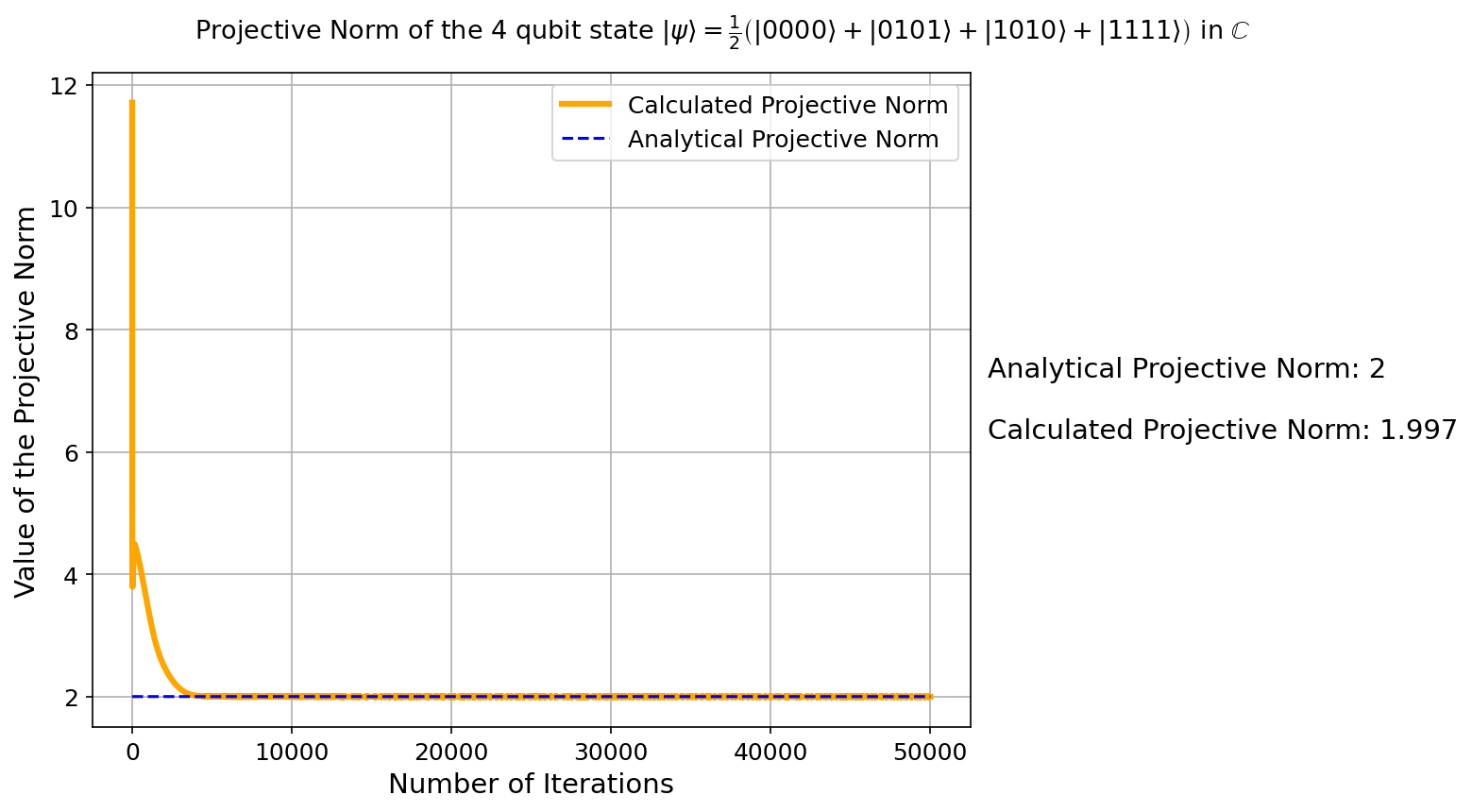}
    \caption{Projective Norm of a 4 qubit state non symmetric state in $\mathbb C$}
    \label{fig:4-qubit-nonsymmetric-complex}
\end{figure}

\begin{figure}[ht]
    \centering
    \includegraphics[width=0.9\linewidth]{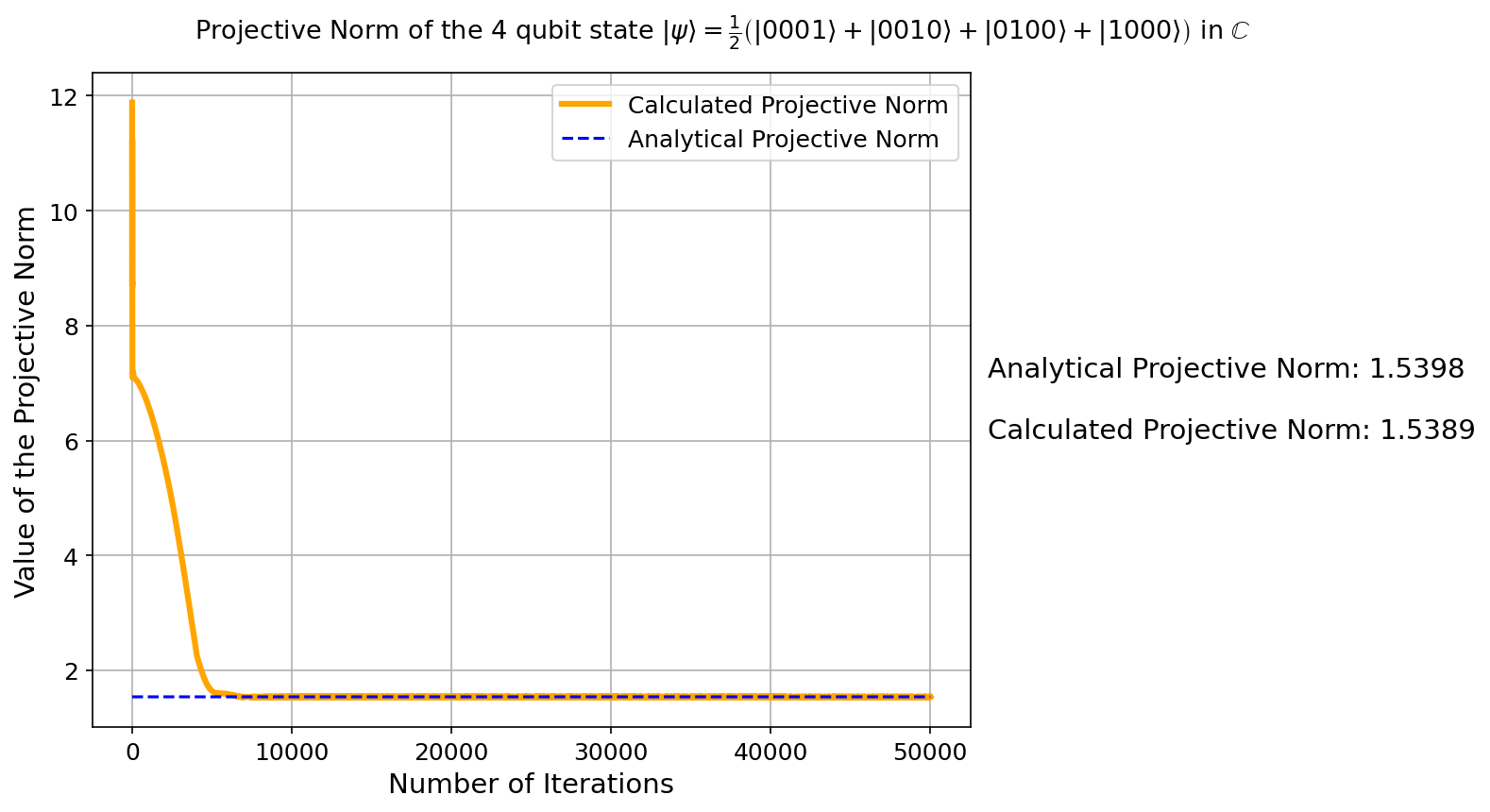}
    \caption{Projective Norm of a 4 qubit state symmetric state in $\mathbb C$}
    \label{fig:4-qubit-symmetric-complex}
\end{figure}

\begin{figure}[ht]
    \centering
    \includegraphics[width=0.97\linewidth]{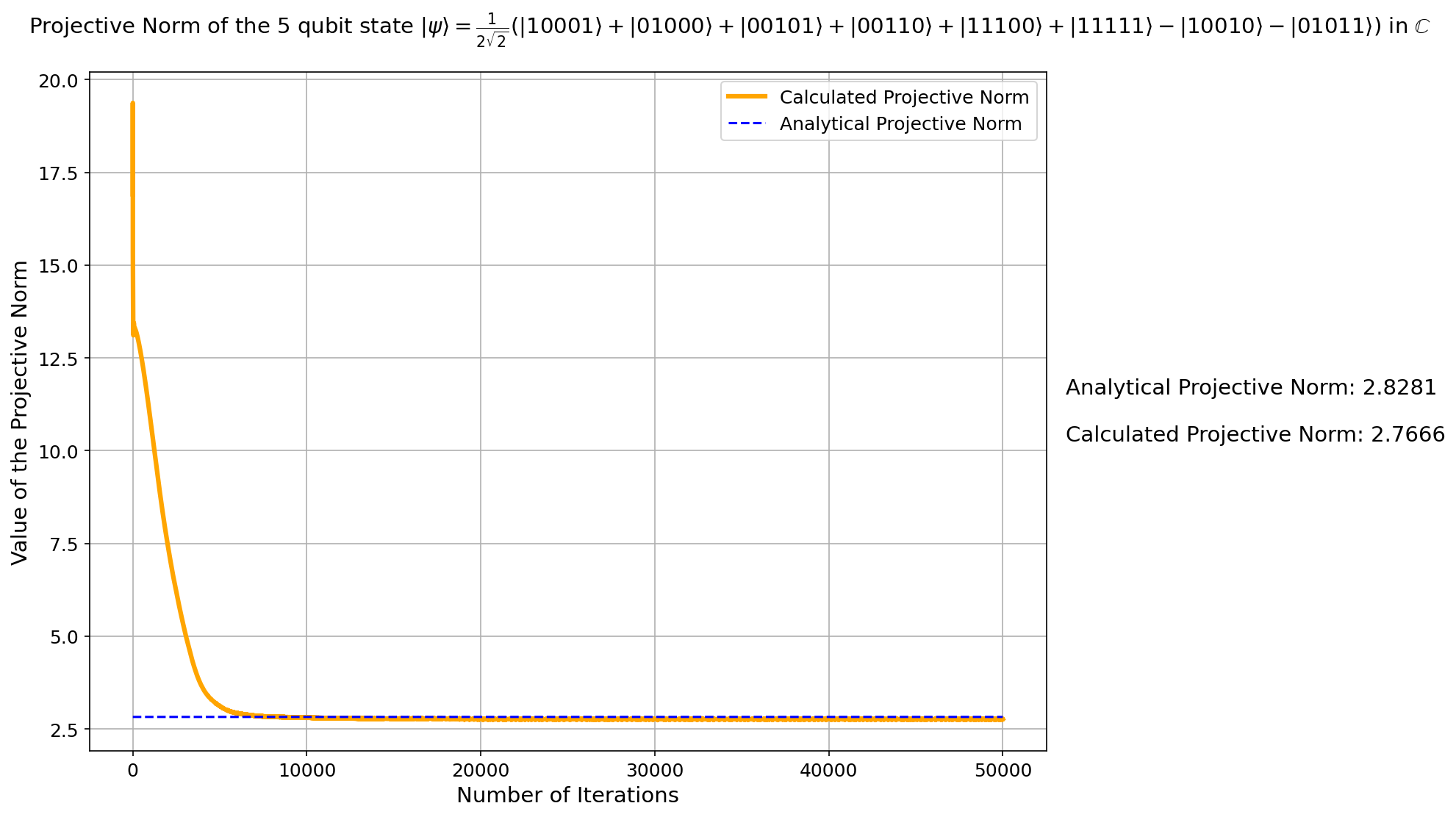}
    \caption{Projective Norm of a 5 qubit state non symmetric state in $\mathbb C$}
    \label{fig:5-qubit-nonsymmetric-complex}
\end{figure}

\begin{figure}[ht]
    \centering
    \includegraphics[width=0.97\linewidth]{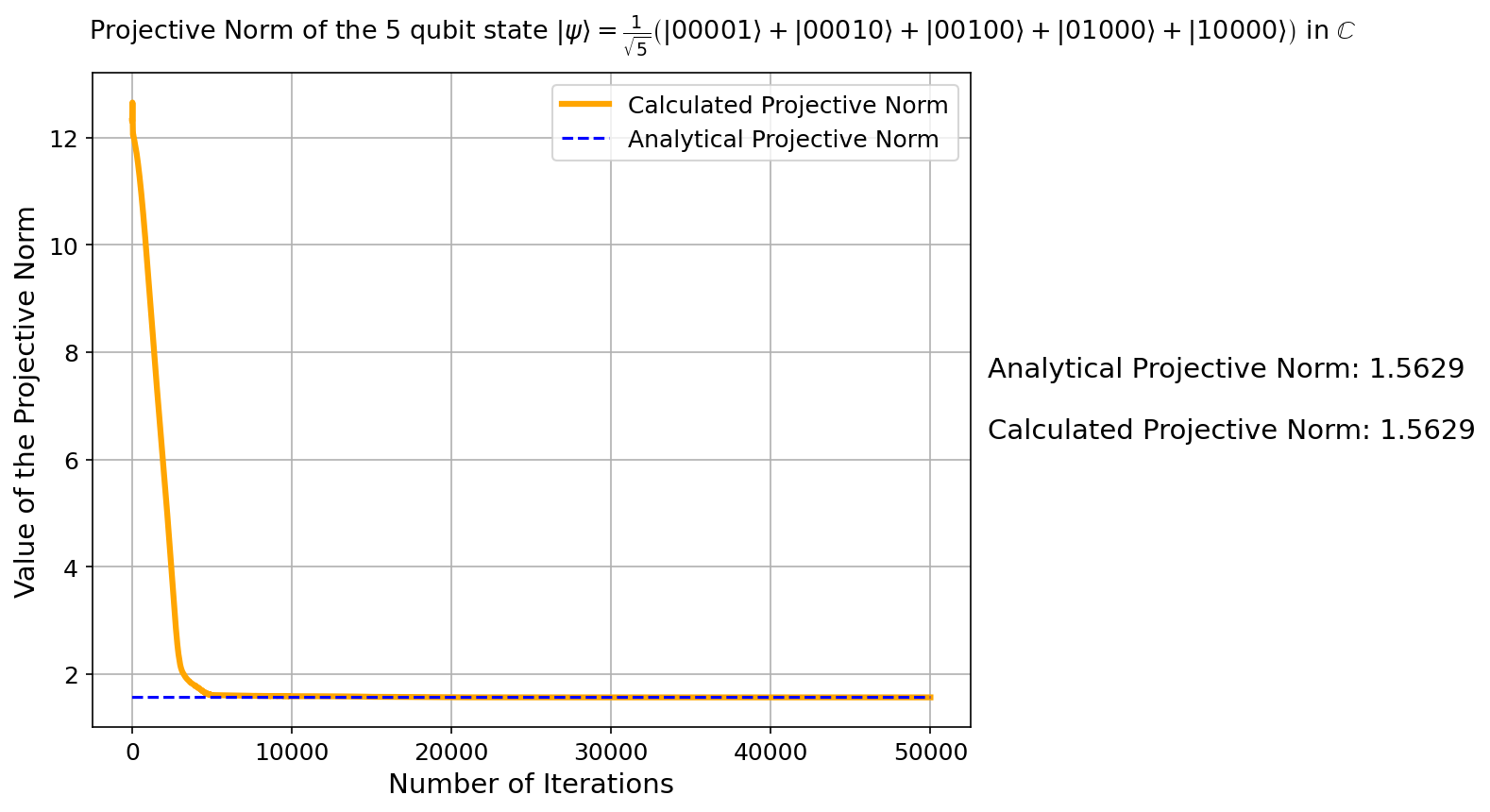}
    \caption{Projective Norm of a 5 qubit state symmetric state in $\mathbb C$}
    \label{fig:5-qubit-symmetric-complex}
\end{figure}

\begin{figure}[ht]
    \centering
    \includegraphics[width=0.97\linewidth]{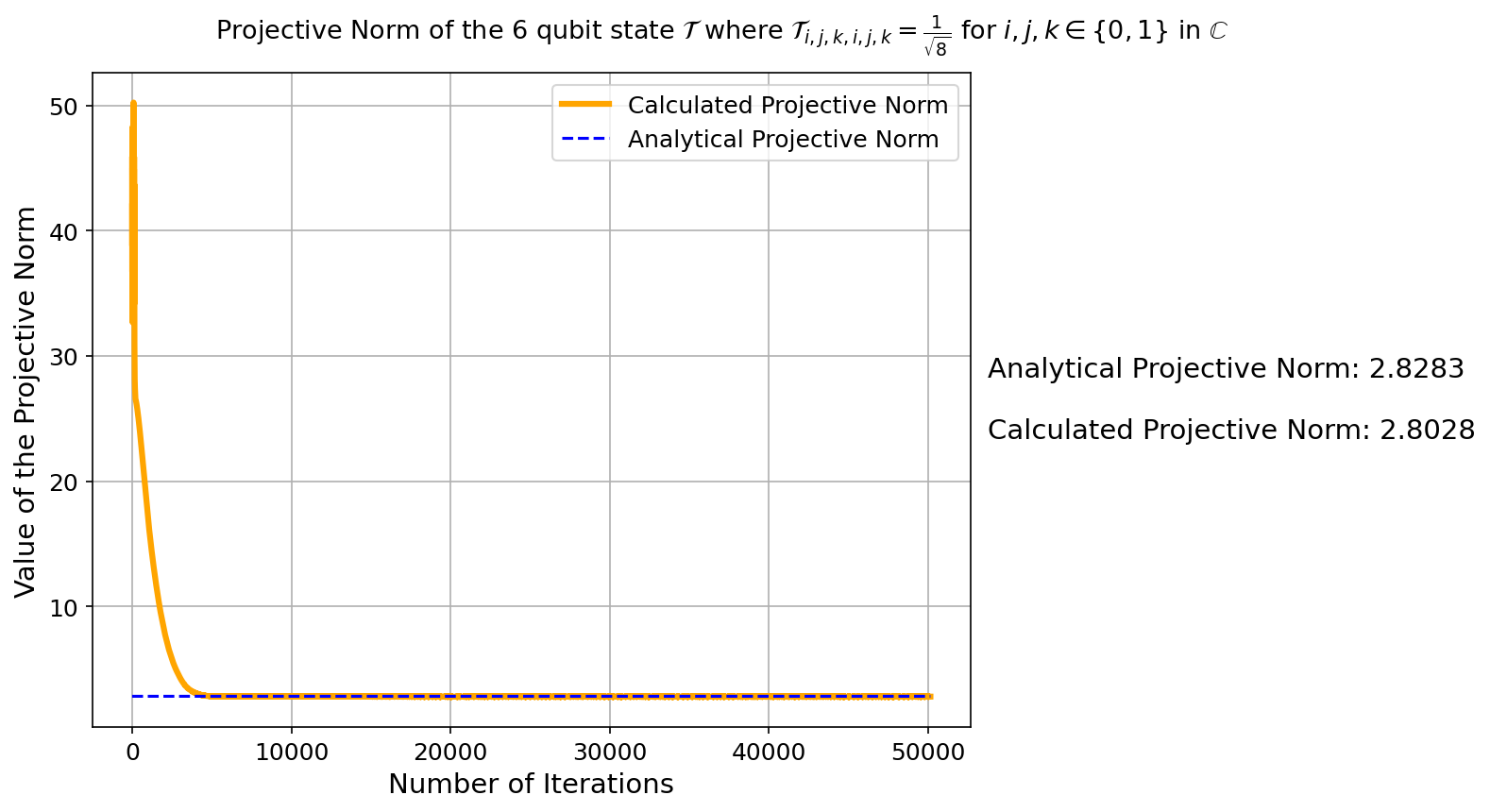}
    \caption{Projective Norm of a 6 qubit state non symmetric state in $\mathbb C$}
    \label{fig:6-qubit-nonsymmetric-complex}
\end{figure}

\begin{figure}[ht]
    \centering
    \includegraphics[width=0.97\linewidth]{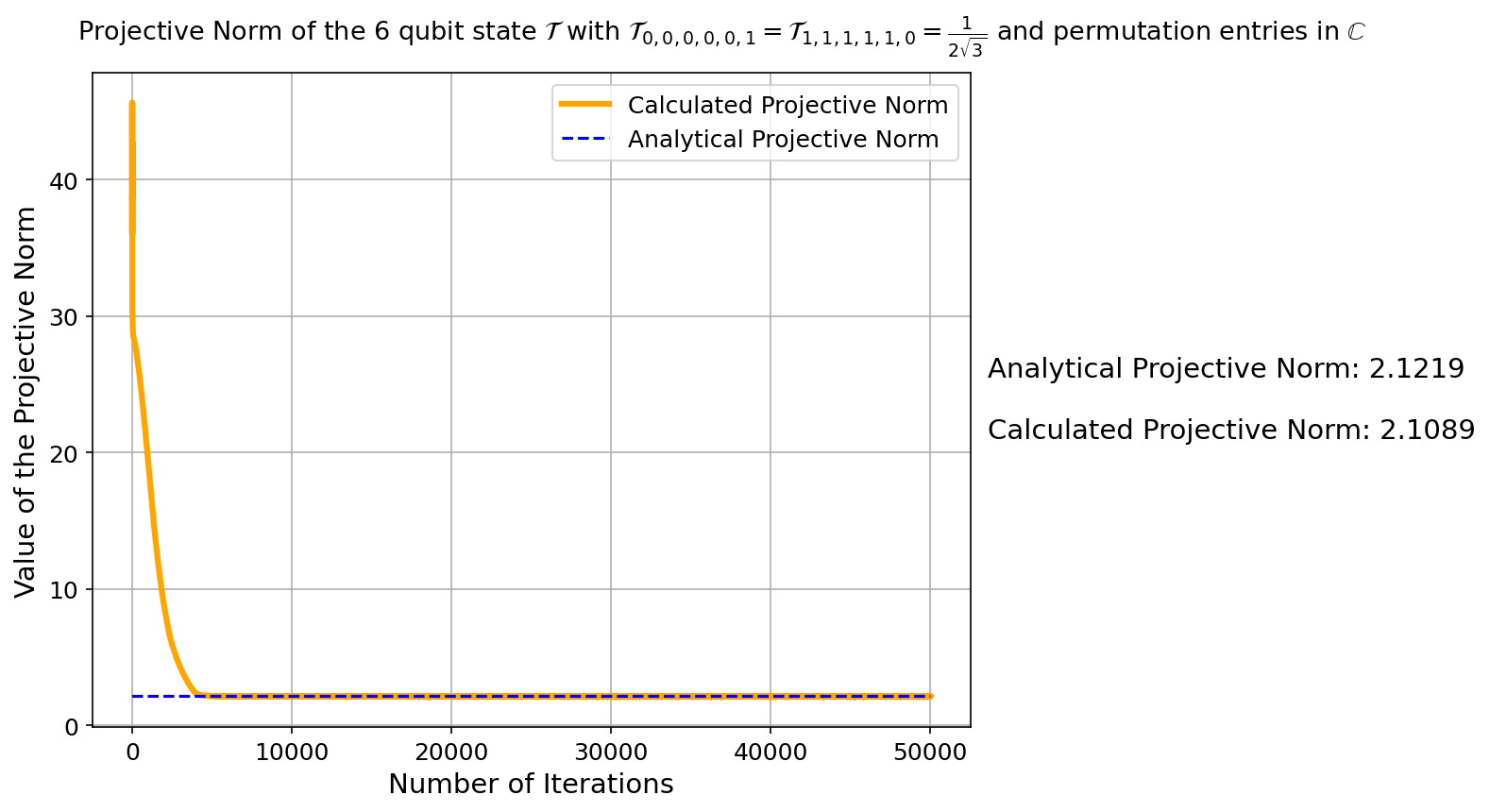}
    \caption{Projective Norm of a 6 qubit state symmetric state in $\mathbb C$}
    \label{fig:6-qubit-symmetric-complex}
\end{figure}

For tensors of order 4 and higher, we do not have analytically calculated results owing to the NP-hard complexity scaling of calculating the projective norm. Hence we utilize the numerical results obtained using the alternating method algorithm in \cite{DFLW17} as our analytical projective norm and compare them with the results obtained using our algorithm. We calculate the projective norm for order 4, 5 and 6 non-symmetric and symmetric tensors to demonstrate the effectiveness of our algorithm for both the general and symmetric respective cases in Fig.\ref{fig:4-qubit-nonsymmetric-complex} and Fig.\ref{fig:4-qubit-symmetric-complex} for order 4 tensors, Fig.\ref{fig:5-qubit-nonsymmetric-complex} and Fig.\ref{fig:5-qubit-symmetric-complex} for order 5 tensors and Fig.\ref{fig:6-qubit-nonsymmetric-complex} and Fig.\ref{fig:6-qubit-symmetric-complex} for order 6 tensors. The time taken for computation increases drastically as the dimensionality of the tensors scales up.

\newpage

\newpage

\newpage

\subsection{Projective Norm Computations for Density Matrices}

We benchmark our algorithm over density matrices by first computing the projective norm over pure state density matrices and then over mixed state density matrices. 

\begin{figure}[ht]
    \centering
    \includegraphics[width= 0.8\linewidth]{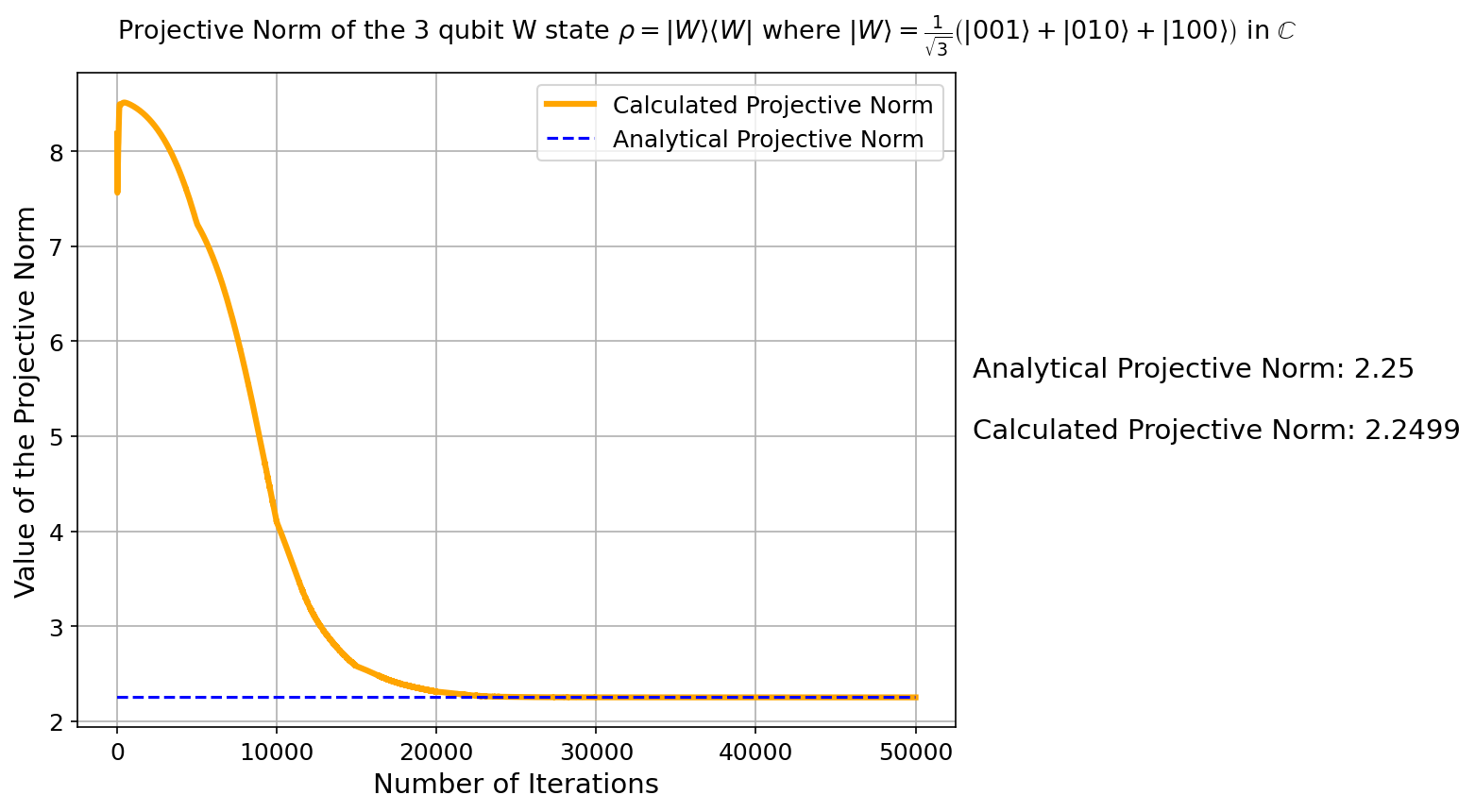}
    \caption{Projective Norm of the 3 qubit W state as a density matrix}
    \label{fig:rho-W-state}
\end{figure}

\begin{figure}[ht]
    \centering
    \includegraphics[width=0.8\linewidth]{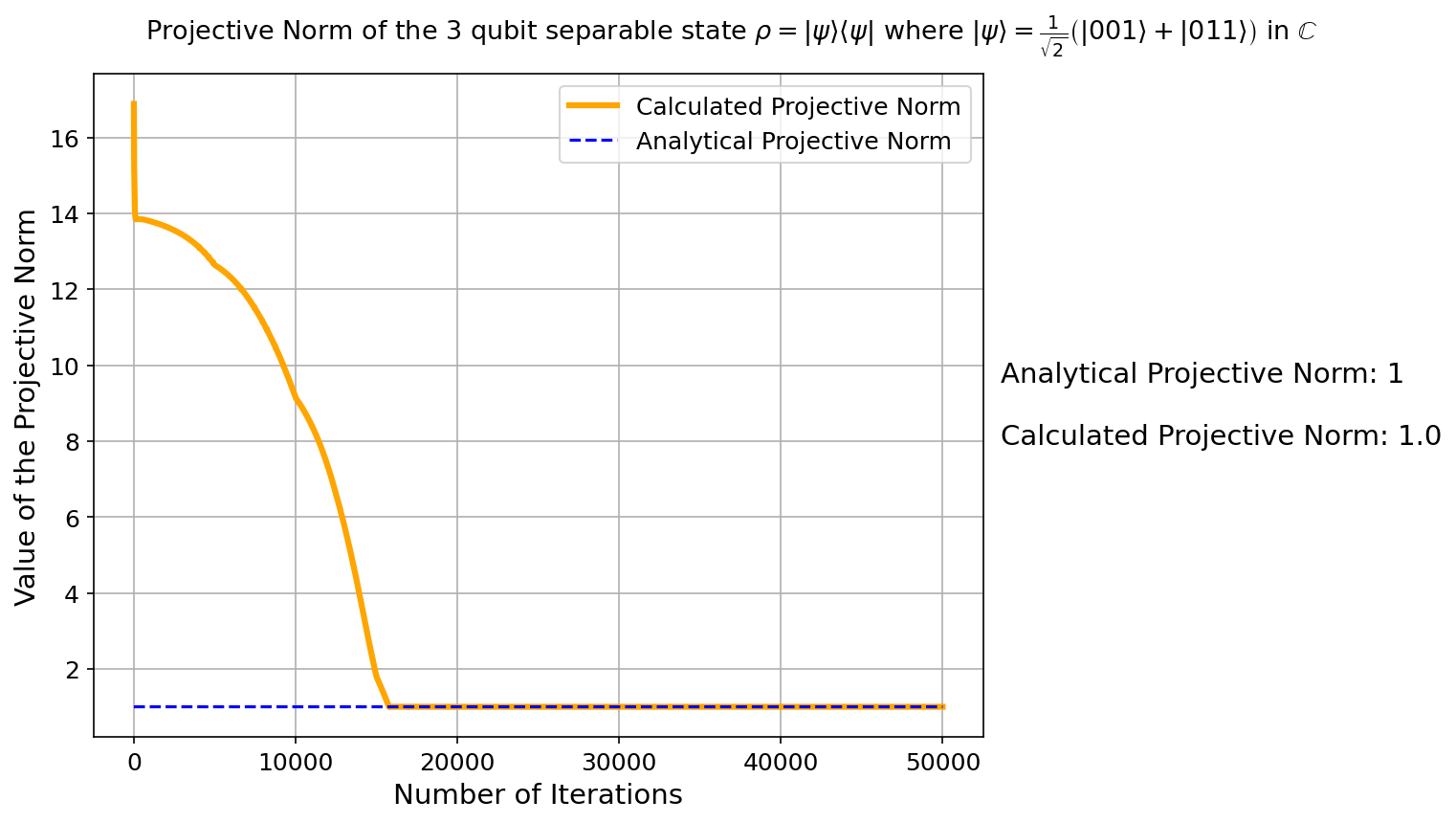}
    \caption{Projective Norm of an arbitrary 3 qubit separable pure state as a density matrix}
    \label{fig:rho-separable-pure-state}
\end{figure}

For the pure state case, we choose the W state for which we already have the analytical projective norm calculated over its tensor form, and an arbitrary 3-qubit separable state for which we know due to its separability that its projective norm is 1. We show the numerical results obtained for these states in Fig.\ref{fig:rho-W-state} and Fig.\ref{fig:rho-separable-pure-state} respectively.

\newpage

For mixed states, we take the following two bipartite state examples given in \cite{DPS04}. These states are parameterized by $\alpha$ over which the state has been found out to be separable or entangled. We verify the same using our projective norm algorithm.

\begin{figure}[ht]
    \centering
    \includegraphics[width=0.9\linewidth]{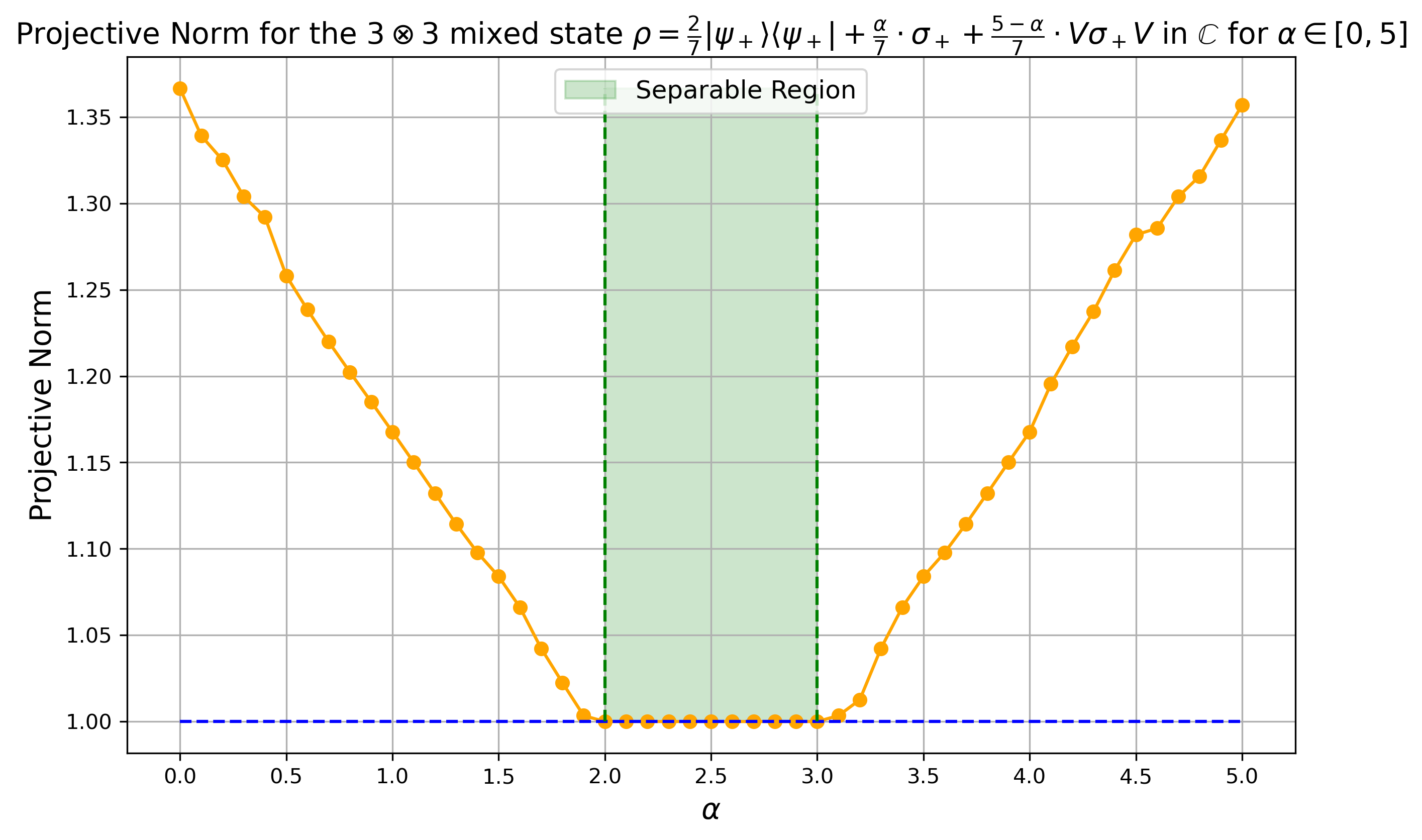}
    \caption{Projective Norm vs $\alpha$ variation for the given $3\otimes3$ bipartite mixed state}
    \label{fig:mixed-state-alpha-norm-eg-1}
\end{figure}

For the $3\otimes3$ state given by:

\begin{equation}
    \rho_\alpha = \frac{2}{7}|\psi_+\rangle\langle\psi_+| + \frac{\alpha}{7}\sigma_+ + \frac{5 - \alpha}{7}V\sigma_+V,
\end{equation}

with $0\le\alpha\le5$, $|\psi_+\rangle = \frac{1}{\sqrt{3}}\sum_{i=0}^2|ii\rangle$, $\sigma_+ = \frac{1}{3}\left(|01\rangle\langle01| + |12\rangle\langle12|+|20\rangle\langle20|\right)$ and $V$ being the operator that swaps the two systems, it is found that the state is separable for $2\le\alpha\le3$ and entangled for the the rest of the values of $\alpha$. We show the same using our algorithm in Fig. \ref{fig:mixed-state-alpha-norm-eg-1} where the projective norm for the mixed state is 1 for $2\le\alpha\le3$ and greater than 1 for the other values of $\alpha$.

\begin{figure}[ht]
    \centering
    \includegraphics[width=0.9\linewidth]{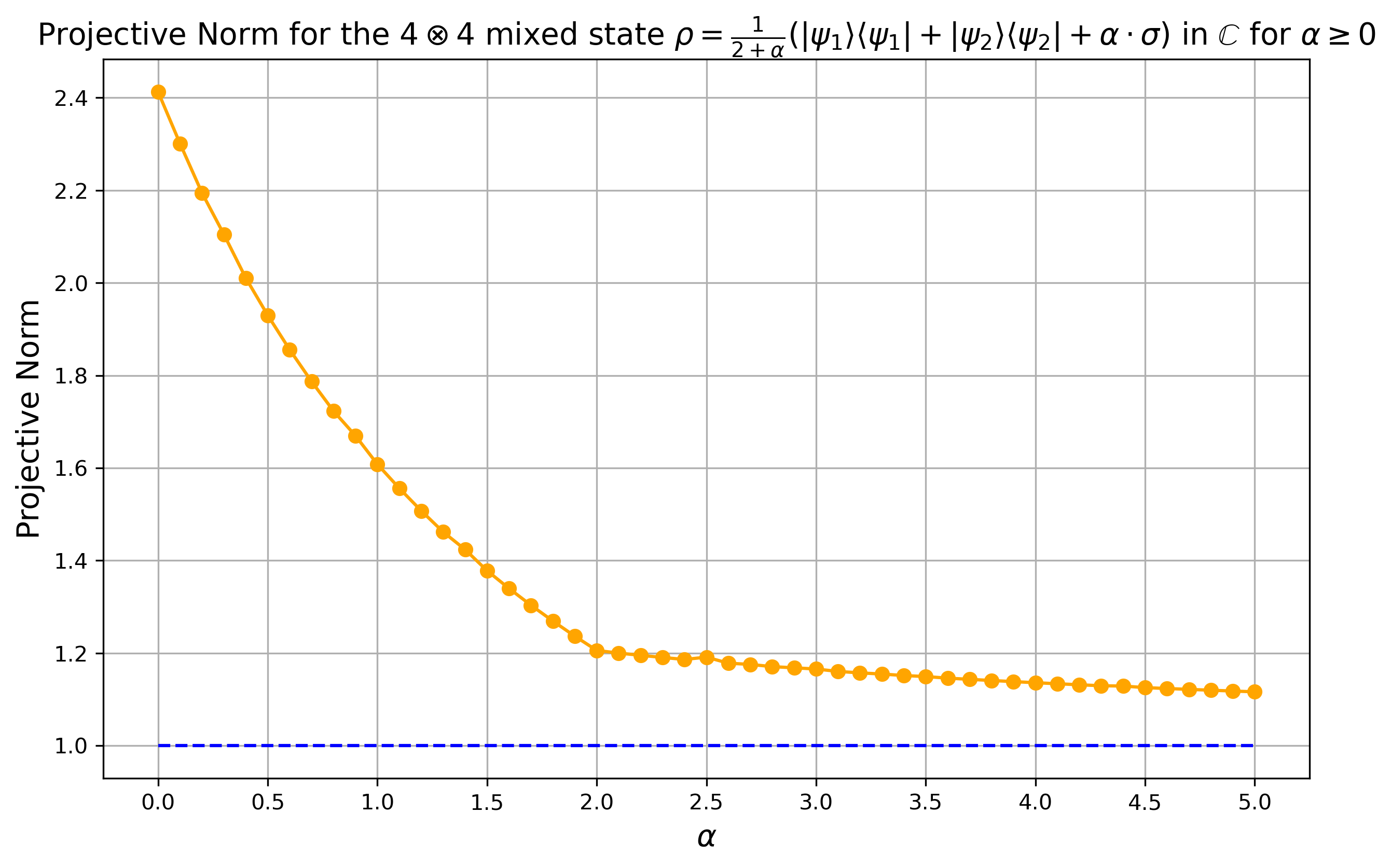}
    \caption{Projective Norm vs $\alpha$ variation for the given $4\otimes4$ bipartite mixed state}
    \label{fig:mixed-state-alpha-norm-eg-2}
\end{figure}

For the $4\otimes4$ state given by:

\begin{equation}
    \rho_\alpha = \frac{1}{2+\alpha}\left(|\psi_1\rangle\langle\psi_1| + |\psi_2\rangle\langle\psi_2| + \alpha\cdot\sigma\right),\text{\hspace{0.25cm}}\alpha\ge 0,
\end{equation}

where
\begin{align*}
    |\psi_1\rangle = &\frac{1}{2}\left(|00\rangle + |11\rangle + \sqrt{2}|22\rangle\right),\\
    |\psi_2\rangle = &\frac{1}{2}\left(|01\rangle + |10\rangle + \sqrt{2}|33\rangle\right),\\
    \sigma = &\frac{1}{8}(|02\rangle\langle02| + |03\rangle\langle03| + |12\rangle\langle12| + |13\rangle\langle13| \\&+ |20\rangle\langle20| + |21\rangle\langle21| + |30\rangle\langle30| + |31\rangle\langle31|),
\end{align*}

it is found that the state is entangled for all values of $\alpha$. We verify the same using our algorithm in Fig. \ref{fig:mixed-state-alpha-norm-eg-2} where the projective norm for this state is observed to be greater than 1 for all values of $\alpha$.

The next example is taken from \cite{ZZZG08} of a $3\otimes3$ state given by

\begin{equation}
    \rho = \frac{1}{8a + 1}
    \begin{pmatrix}
        a & 0 & 0 & 0 & a & 0 & 0 & 0 & a \\
        0 & a & 0 & 0 & 0 & 0 & 0 & 0 & 0 \\
        0 & 0 & a & 0 & 0 & 0 & 0 & 0 & 0 \\
        0 & 0 & 0 & a & 0 & 0 & 0 & 0 & 0 \\
        a & 0 & 0 & 0 & a & 0 & 0 & 0 & a \\
        0 & 0 & 0 & 0 & 0 & a & 0 & 0 & 0 \\
        0 & 0 & 0 & 0 & 0 & 0 & \frac{1 + a}{2} & 0 & \frac{\sqrt{1 - a^2}}{2} \\
        0 & 0 & 0 & 0 & 0 & 0 & 0 & a & 0 \\
        a & 0 & 0 & 0 & a & 0 & \frac{\sqrt{1 - a^2}}{2} & 0 & \frac{1 + a}{2}
    \end{pmatrix}
\end{equation}

where $0< a < 1$. We consider a mixture of this state with white noise

\begin{equation}
    \rho(a, p) = p\cdot\rho(a)  + (1 - p)\cdot \frac{\mathbb{I}}{9}, \text{\hspace{0.25cm}} 0\le p\le 1.
\end{equation}

\begin{figure}[ht]
    \centering
    \includegraphics[width=\linewidth]{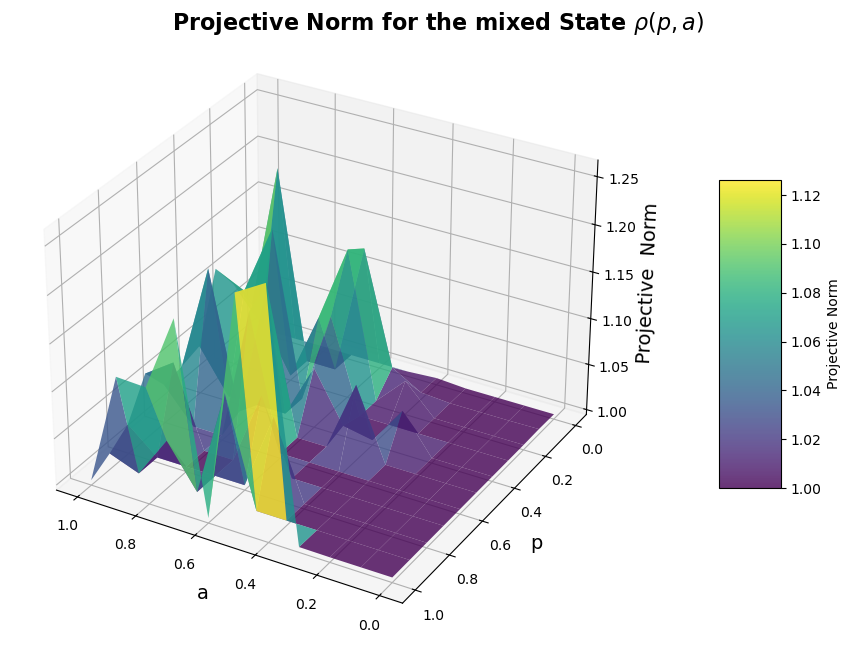}
    \caption{Projective Norm variation for different values of a and p for the given $3\otimes3$ bipartite mixed state}
    \label{fig:mixed-state-rho-a-p}
\end{figure}

\begin{figure}[ht]
    \centering
    \includegraphics[width=0.8\linewidth]{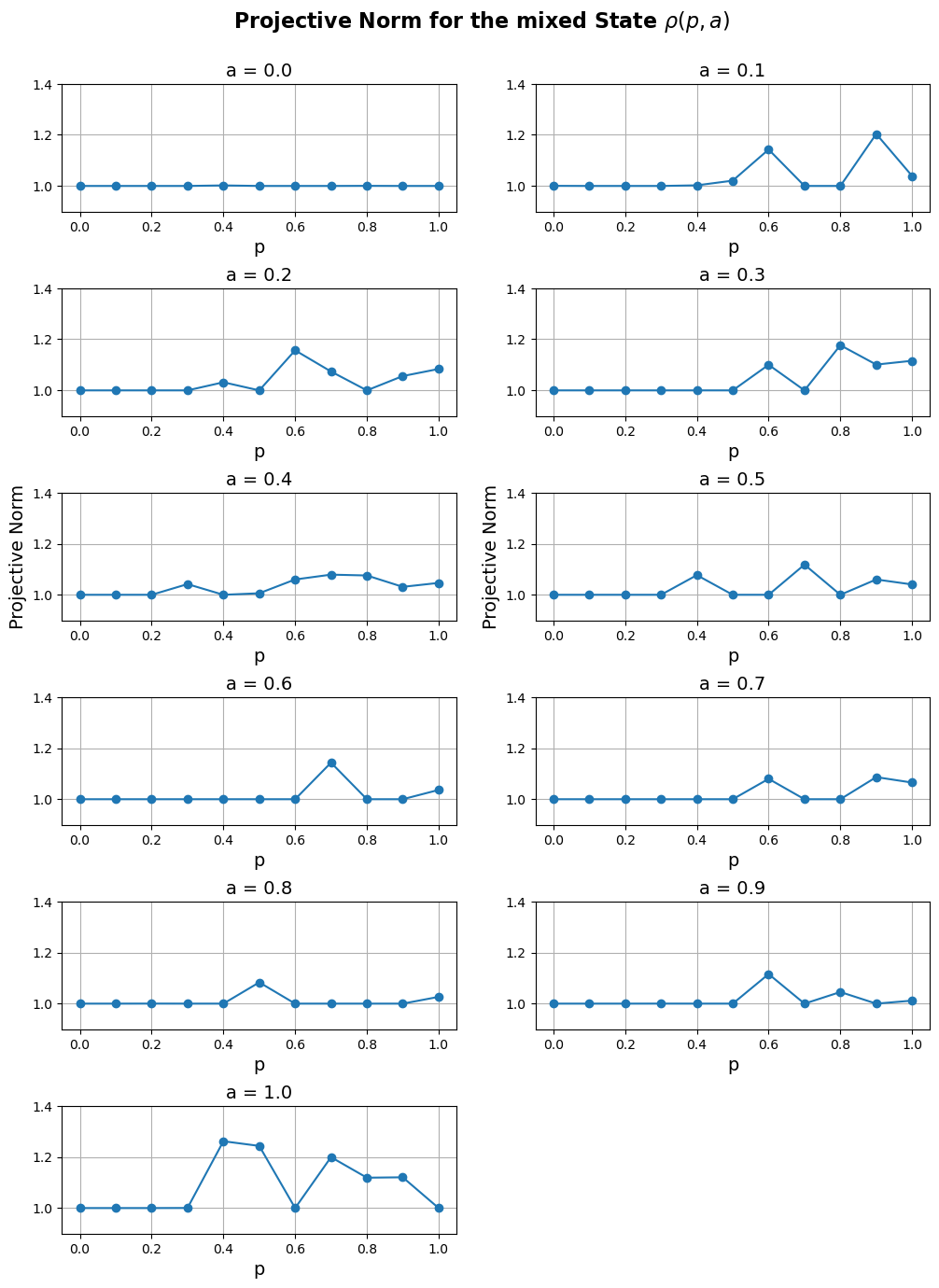}
    \caption{Projective Norm variation for different values of a for the given $3\otimes3$ bipartite mixed state}
    \label{fig:mixed-state-rho-a-p-fixed-a}
\end{figure}

Fig. \ref{fig:mixed-state-rho-a-p} and Fig. \ref{fig:mixed-state-rho-a-p-fixed-a} show the variation of the projective norm with respect to the parameters $a$ and $p$ showcasing the degree of entanglement for different values of the parameters.

\newpage

\newpage

\newpage

\section{Conclusions}

The projective tensor norm is an important tool for detecting and quantifying entanglement for a given multipartite quantum state. It can even be utilized to measure entanglement for quantum states having order greater than two, therefore providing a more generalized method to detect entanglement than methods such as positive trace maps, etc. However, calculating projective norms is an NP-hard problem computationally. Existing algorithms do not guarantee convergence and are restricted to pure states. As discussed in this paper, we present a novel gradient descent-based algorithm for computing the projective tensor norm. Our developed algorithm guarantees convergence to the global minimum norm value, i.e. the projective norm. We demonstrated this convergence for higher order tensors using previously calculated analytical and numerical values, showing the effectiveness of our algorithm in tackling quantum states from large Hilbert spaces. We also show that our algorithm can be extended to calculate the projective norm for density matrices of both pure and mixed states. We explicitly showcase the performance of our algorithm for mixed states with parameterized introduction of noise, displaying the sensitivity of the algorithm for detecting entanglement in noisy states. The code is available open source on GitHub \cite{RJ25}.

\vspace{1em}

\noindent{\textit{Acknowledgements.} A.R. and M.A.J. thank Ion Nechita and Khurshed Fitter for their invaluable discussions in understanding projective norms questions. A.R. also thanks Stuti Pradhan, Aditya Jivoji and Tabitha Sneha for their technical support in running the experiments.}

\addcontentsline{toc}{section}{References}

$\,$

$\,$

\end{document}